\documentclass[zpreprint,zbstdefault]{zeus_paper}
\usepackage{subfigure}
\usepackage[english]{babel}

\newcommand{\ZcoosysB}{%
The ZEUS coordinate system is a right-handed Cartesian system, with the $Z$
axis pointing in the proton beam direction, referred to as the ``forward
direction'', and the $X$ axis pointing left towards the center of HERA.
The coordinate origin is at the nominal interaction point.\xspace}
\newcommand{\Zpsrap}{%
The pseudorapidity is defined as $\eta=-\ln\left(\tan\frac{\theta}{2}\right)$,
where the polar angle, $\theta$, is measured with respect to the proton beam
direction.\xspace}

\newcommand{\ZcoosysfnBeta}{\footnote{\ZcoosysB\Zpsrap}}
\newcommand{\Zctddesc}[1]{%
Charged particles are tracked in the central tracking detector (CTD)~\citeCTD,
which operates in a magnetic field of $1.43\Tesla$ provided by a thin 
superconducting coil. The CTD consists of 72~cylindrical drift chamber 
layers, organized in 9~superlayers covering the polar-angle#1 region 
\mbox{$15^\circ<\theta<164^\circ$}. The transverse-momentum resolution for
full-length tracks is $\sigma(p_T)/p_T=0.0058p_T\oplus0.0065\oplus0.0014/p_T$,
with $p_T$ in $\Gev$.}
\newcommand{\Zcaldesc}{%
The high-resolution uranium--scintillator calorimeter (CAL)~\citeCAL consists 
of three parts: the forward (FCAL), the barrel (BCAL) and the rear (RCAL)
calorimeters. Each part is subdivided transversely into towers and
longitudinally into one electromagnetic section (EMC) and either one (in RCAL)
or two (in BCAL and FCAL) hadronic sections (HAC). The smallest subdivision of
the calorimeter is called a cell.  The CAL energy resolutions, as measured under
test-beam conditions, are $\sigma(E)/E=0.18/\sqrt{E}$ for electrons and
$\sigma(E)/E=0.35/\sqrt{E}$ for hadrons, with $E$ in $\Gev$.}
\chardef\usc=95
\chardef\til=126
\catcode`\@=11 % @ signs are now treated as letters
\DeclareRobustCommand\xdotspace{\futurelet\@let@token\@xdotspace}
\def\@xdotspace{%
  \ifx\@let@token.\else
  \ifx\@let@token\bgroup.\else
  \ifx\@let@token\egroup.\else
  \ifx\@let@token\/.\else
  \ifx\@let@token\ .\else
  \ifx\@let@token~.\else
  \ifx\@let@token!.\else
  \ifx\@let@token,.\else
  \ifx\@let@token:.\else
  \ifx\@let@token;.\else
  \ifx\@let@token?.\else
  \ifx\@let@token/.\else
  \ifx\@let@token'.\else
  \ifx\@let@token).\else
  \ifx\@let@token-.\else
  \ifx\@let@token\@xobeysp.\else
  \ifx\@let@token\space.\else
  \ifx\@let@token\@sptoken.\else
   .\space
   \fi\fi\fi\fi\fi\fi\fi\fi\fi\fi\fi\fi\fi\fi\fi\fi\fi\fi}
\catcode`\@=12 % @ signs are no longer letters

\newcommand{\stru}[2]{%
   \relax\ifmmode\hbox{\vrule height#1 depth#2 width0pt}%
   \else\vrule height#1 depth#2 width0pt\fi}
\newcommand{\Ronum}[1]{\uppercase\expandafter{\romannumeral#1}}
\newcommand{\ronum}[1]{\expandafter{\romannumeral#1}}
\DeclareRobustCommand{\LaTeXZ}{%
  \LaTeX\kern-.05em4\kern-.1em
  {\raisebox{-0.2ex}{$\scriptstyle\text{ZEUS}$}}\xspace}
\DeclareMathAlphabet{\mathbf}{OT1}{cmr}{bx}{sl}
\newcommand{\eVdist}{\kern-0.06667em}

\newcommand{\Gev}{{\text{Ge}\eVdist\text{V\/}}}

\newcommand{\Tesla}{\,\text{T}}

\newcommand{\slashfrac}[2]{%
  \raisebox{0.5ex}{\ensuremath #1}\kern-0.12em/\kern-0.08em
  \raisebox{-.8ex}{\ensuremath #2}}
\newcommand{\sqr}[3]{%
    {\vcenter{\hrule height.#3ex\hbox{\vrule width.#2ex height#1ex
     \kern#1ex\vrule width.#3ex}\hrule height.#2ex}}}

\catcode`\@=11 % @ signs are now treated as letters
\newcommand{\parenbar}{\mathpalette\p@renb@r}
\def\p@renb@r#1#2{\vbox{%
  \ifx#1\scriptscriptstyle \dimen@.7em\dimen@ii.2em\else
  \ifx#1\scriptstyle \dimen@.8em\dimen@ii.25em\else
  \dimen@1em\dimen@ii.4em\fi\fi \offinterlineskip
  \ialign{\hfill##\hfill\cr
    \vbox{\hrule width\dimen@ii}\cr
    \noalign{\vskip-.3ex}%
    \hbox to\dimen@{$\mathchar300\hfil\mathchar301$}\cr
    \noalign{\vskip-.3ex}%
    $#1#2$\cr}}}
\catcode`\@=12 % @ signs are no longer letters

\newcommand{\IP}{{\rm I$\kern-0.01667em$P}\xspace}

\mathchardef\qsm=63
\mathchardef\pls=43
\mathchardef\mns=512
\mathchardef\plm=518
\mathchardef\eql=61
\mathchardef\smallleft=300
\mathchardef\smallright=301
\mathchardef\les=316
\mathchardef\gre=318
\mathchardef\leq=532
\mathchardef\grq=533
\catcode`\@=11 % @ signs are now treated as letters
\newcounter{pict@width}
\newcounter{pict@height}
\newlength{\pict@scale}
\newcommand{\psfigadd}[4]{%
\setcounter{pict@width}{1*\ratio{#2+\pict@scale/2}{\pict@scale}}
\setcounter{pict@height}{1*\ratio{#3+\pict@scale/2}{\pict@scale}}
\setlength{\unitlength}{\pict@scale}
\hbox to #2{\hspace{-\fill}\begin{picture}(\thepict@width,\thepict@height)
\put(0,0){\psfig{figure=#1,width=#2,height=#3,clip=}}
\SetScale{0.283466457}
\SetWidth{1.763889}
{#4}
\end{picture}}
}
\newcounter{pict@widthfst}
\newcounter{pict@widthscd}
\newcounter{pict@widthtot}
\newcommand{\psfigaddtwo}[7]{%
\setcounter{pict@widthfst}{1*\ratio{#2+\pict@scale/2}{\pict@scale}}
\setcounter{pict@widthscd}{1*\ratio{#2+#4+\pict@scale/2}{\pict@scale}}
\setcounter{pict@widthtot}{1*\ratio{#2+#4+#6+\pict@scale/2}{\pict@scale}}
\setcounter{pict@height}{1*\ratio{#3+\pict@scale/2}{\pict@scale}}
\setlength{\unitlength}{\pict@scale}
\hbox{\hspace{-\fill}\begin{picture}(\thepict@widthtot,\thepict@height)
\put(0,0){\psfig{figure=#1,width=#2,height=#3,clip=}}
\put(\thepict@widthscd,0){\psfig{figure=#5,width=#6,height=#3,clip=}}
\SetScale{0.283466457}
\SetWidth{1.763889}
{#7}
\end{picture}}
}
\newcommand{\psfigror}[4]{%
\setcounter{pict@width}{1*\ratio{#2+\pict@scale/2}{\pict@scale}}
\setcounter{pict@height}{1*\ratio{#3+\pict@scale/2}{\pict@scale}}
\setlength{\unitlength}{\pict@scale}
\hbox{\begin{picture}(\thepict@width,\thepict@height)
\put(0,\thepict@height){\psfig{figure=#1,width=#3,height=#2,clip=,angle=270}}
\SetScale{0.283466457}
\SetWidth{1.763889}
{#4}
\end{picture}}
}
\newcommand{\psfigrol}[4]{%
\setcounter{pict@width}{1*\ratio{#2+\pict@scale/2}{\pict@scale}}
\setcounter{pict@height}{1*\ratio{#3+\pict@scale/2}{\pict@scale}}
\setlength{\unitlength}{\pict@scale}
\hbox{\begin{picture}(\thepict@width,\thepict@height)
\put(0,0){\psfig{figure=#1,width=#3,height=#2,clip=,angle=90}}
\SetScale{0.283466457}
\SetWidth{1.763889}
{#4}
\end{picture}}
}
\catcode`\@=12 % @ signs are no longer letters
\newlength\listtextwidth

\catcode`\@=11 % @ signs are now treated as letters
\newlength{\@tabfninsert}
\newlength{\@tabfnwidth}
\newcommand{\tabfootnote}[2]{%
  \setlength{\@tabfninsert}{0.8em}
  \setlength{\@tabfnwidth}{\textwidth}
  \addtolength{\@tabfnwidth}{-\@tabfninsert}
  \addtolength{\@tabfnwidth}{-0.4em}
  \noindent\makebox[\@tabfninsert][r]{\footnotesize$^{#1}$\hfil}\hfill%
  \parbox[t]{\@tabfnwidth}{\footnotesize #2\hfill}}
\catcode`\@=12 % @ signs are no longer letters

\newcommand{\xgo} {\mbox{$x_{\gamma}^{\rm obs}$}}

\def\3{\ss}

\def\citeCTD{{\cite{%
nim:a279:290,*npps:b32:181,*nim:a338:254%
}}\xspace}
\def\citeCAL{{\cite{%
nim:a309:77,*nim:a309:101,*nim:a321:356,*nim:a336:23%
}}\xspace}

\begin{document}
%------------------------------------------------------------------------------
%       Title sheet
%------------------------------------------------------------------------------
\title{
\vspace{-5cm}
\begin{flushright} {\normalsize \tt DESY 07-092}\\ \vspace{-.25cm}{\normalsize \tt June 2007} \end{flushright}
\vspace{2cm}
High-$E_{T}$ dijet photoproduction at HERA\\
}                                                       
                    
\author{ZEUS Collaboration}
\draftversion{}
%\date{3\ May 2007}

\abstract{
The cross section for high-$E_{T}$ dijet production in photoproduction has been 
measured with the ZEUS detector at HERA using an integrated luminosity of 
$81.8$\,pb$^{-1}$. The events were required to have a virtuality of the incoming 
photon, $Q^{2}$, of less than $1$\,GeV$^{2}$ and a photon-proton center-of-mass 
energy in the range $142 < W_{\gamma p} < 293$\,GeV. Events were selected if at 
least two jets satisfied the  transverse-energy requirements of 
$E_{T}^{\rm jet1} > 20$\,GeV and $E_{T}^{\rm jet2} > 15$\,GeV and pseudorapidity 
(with respect to the proton beam direction) 
requirements of $-1 < \eta^{\rm jet1,2} < 3$, with at least one of the jets 
satisfying $-1 < \eta^{\rm jet} < 2.5$. The measurements show sensitivity to 
the parton distributions in the photon and proton and to effects beyond next-to-leading 
order in QCD. Hence these data can be used to constrain further the parton densities 
in the proton and photon. 
}

\makezeustitle

\def\3{\ss}                                                                                        
\pagenumbering{Roman}                                                                              
                                    % this "%"s are for cosmetics only                             
% \begin{document}                                                                                   
                                                   %                                               
\begin{center}                                                                                     
{                      \Large  The ZEUS Collaboration              }                               
\end{center}                                                                                       
  S.~Chekanov$^{   1}$,                                                                            
  M.~Derrick,                                                                                      
  S.~Magill,                                                                                       
  B.~Musgrave,                                                                                     
  D.~Nicholass$^{   2}$,                                                                           
  \mbox{J.~Repond},                                                                                
  R.~Yoshida\\                                                                                     
 {\it Argonne National Laboratory, Argonne, Illinois 60439-4815}, USA~$^{n}$                       
\par \filbreak                                                                                     
  M.C.K.~Mattingly \\                                                                              
 {\it Andrews University, Berrien Springs, Michigan 49104-0380}, USA                               
\par \filbreak                                                                                     
  M.~Jechow, N.~Pavel~$^{\dagger}$, A.G.~Yag\"ues Molina \\                                        
  {\it Institut f\"ur Physik der Humboldt-Universit\"at zu Berlin,                                 
           Berlin, Germany}
% ...
\newpage                                                                                           
$^{\    1}$ supported by DESY, Germany \\                                                          
$^{\    2}$ also affiliated with University College London, UK \\                                  
% ...
$^{\dagger}$ deceased \\                                                                           
%                                                                                                  
% \par         % if index listing & table fit to 1 page, put gap here                              
%\newpage   % alternatively: go to newpage, if page is too small                                    
                                                           %                                       
% \institute_references_start    % do not touch or move this line !                                
                                                           %                                       
\begin{tabular}[h]{rp{14cm}}                                                                       
$^{a}$ &  supported by the Natural Sciences and Engineering Research Council of Canada (NSERC) \\  
$^{b}$ &  supported by the German Federal Ministry for Education and Research (BMBF), under        
          contract numbers HZ1GUA 2, HZ1GUB 0, HZ1PDA 5, HZ1VFA 5\\                                
$^{c}$ &  supported in part by the MINERVA Gesellschaft f\"ur Forschung GmbH, the Israel Science   
          Foundation (grant no. 293/02-11.2) and the U.S.-Israel Binational Science Foundation \\  
$^{d}$ &  supported by the German-Israeli Foundation and the Israel Science Foundation\\           
$^{e}$ &  supported by the Italian National Institute for Nuclear Physics (INFN) \\                
$^{f}$ &  supported by the Japanese Ministry of Education, Culture, Sports, Science and Technology 
          (MEXT) and its grants for Scientific Research\\                                          
$^{g}$ &  supported by the Korean Ministry of Education and Korea Science and Engineering          
          Foundation\\                                                                             
$^{h}$ &  supported by the Netherlands Foundation for Research on Matter (FOM)\\                   
$^{i}$ &  supported by the Polish State Committee for Scientific Research, grant no.               
          620/E-77/SPB/DESY/P-03/DZ 117/2003-2005 and grant no. 1P03B07427/2004-2006\\             
$^{j}$ &  partially supported by the German Federal Ministry for Education and Research (BMBF)\\   
$^{k}$ &  supported by RF Presidential grant N 8122.2006.2 for the leading                         
          scientific schools and by the Russian Ministry of Education and Science through its grant
          Research on High Energy Physics\\                                                        
$^{l}$ &  supported by the Spanish Ministry of Education and Science through funds provided by     
          CICYT\\                                                                                  
$^{m}$ &  supported by the Particle Physics and Astronomy Research Council, UK\\                   
$^{n}$ &  supported by the US Department of Energy\\                                               
$^{o}$ &  supported by the US National Science Foundation. Any opinion,                            
findings and conclusions or recommendations expressed in this material                             
are those of the authors and do not necessarily reflect the views of the                           
National Science Foundation.\\                                                                     
$^{p}$ &  supported by the Polish Ministry of Science and Higher Education                         
as a scientific project (2006-2008)\\                                                              
$^{q}$ &  supported by FNRS and its associated funds (IISN and FRIA) and by an Inter-University    
          Attraction Poles Programme subsidised by the Belgian Federal Science Policy Office\\     
$^{r}$ &  supported by the Malaysian Ministry of Science, Technology and                           
Innovation/Akademi Sains Malaysia grant SAGA 66-02-03-0048\\                                       
\end{tabular}                                                                                      
                                                           %                                       
% \institute_references_end     % do not touch or move this line !                                 
                                                           %                                       
%\end{document}                                                                                     

%------------------------------------------------------------------------------
%       Text
%------------------------------------------------------------------------------
\pagenumbering{arabic} 
\pagestyle{plain}
% ----------------------------------------------------------------------------
%       Introduction
% ----------------------------------------------------------------------------
\section{Introduction}
\label{sec-int}

\begin{figure}[htb]
\begin{center}
~\epsfig{file=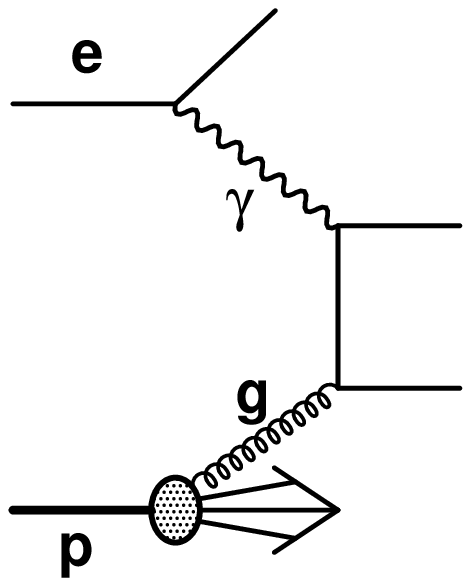,height=6cm}
\hspace{1cm}
~\epsfig{file=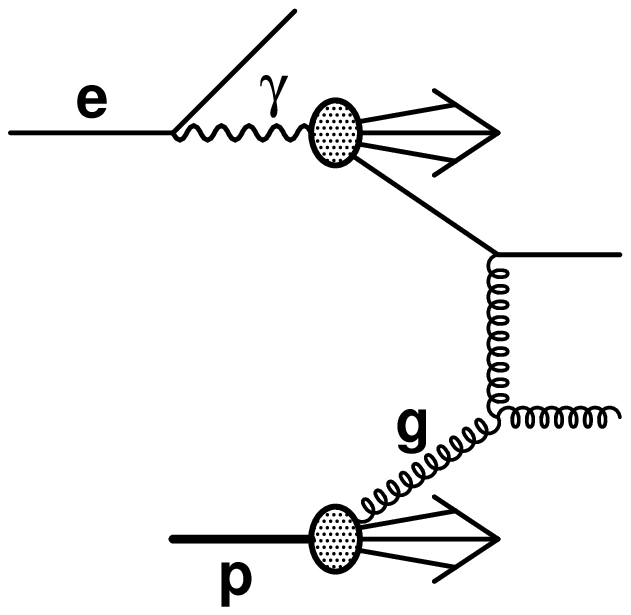,height=6cm}
\put(-298,0){\makebox(0,0)[tl]{\large (a)}}
\put(-88,0){\makebox(0,0)[tl]{\large (b)}}
\vspace{1cm}
\caption{Examples of (a) direct and (b) resolved dijet photoproduction diagrams in 
positron-proton, $ep$, collisions in LO QCD. This direct-photon process is the collision 
of a photon, $\gamma$, and gluon, $g$ from the proton. This resolved-photon process is a 
collision of a parton from the photon and a gluon, $g$, from the proton.}
\label{fig:feyn}
\end{center}
\end{figure}

In photoproduction at HERA, a quasi-real photon emitted from the incoming 
positron\footnote{In the following, the term ``positron'' denotes generically 
both the electron ($e^-$) and positron ($e^+$). Unless explicitly stated, 
positron will be the term used to describe both particles.} collides with 
a parton from the incoming proton. The photoproduction of jets can be 
classified into two types of processes in leading-order (LO) Quantum 
Chromodynamics (QCD). In direct processes, the photon participates in the hard 
scatter via either boson-gluon fusion (see Fig.~\ref{fig:feyn}(a)) or QCD Compton 
scattering. The second class, resolved processes (see Fig.~\ref{fig:feyn}(b)), 
involves the photon acting as a source of quarks 
and gluons, with only a fraction of its momentum, $x_\gamma$, participating in the hard 
scatter. Measurements of jet cross sections in 
photoproduction~\cite{pl:b322:287,epj:c1:109,pl:b531:9,epj:c23:615,pl:b342:417,*pl:b348:665,*pl:b384:401,*pl:b443:394,*epj:c4:591,*epj:c11:35,*pl:b560:7,pl:b314:436,*zfp:c70:17,*epj:c1:97,*pl:b483:36,*epj:c25:13,*epj:c29:497,*pl:b639:21} are sensitive to the structure 
of both the proton and the photon and thus provide input to global fits to determine 
their parton densities.

There are three objectives of the measurement reported in this paper. Firstly, the 
analysis was designed to provide constraints on the parton density functions (PDFs) 
of the photon. Over the last two years there has been active research in the 
area of fitting photon PDFs and a number of new parameterizations 
have become available~\cite{pr:d70:093004,epj:c34:395,epj:c45:633}. In two 
of these~\cite{pr:d70:093004,epj:c34:395}, fits were performed exclusively to photon 
structure function, $F_{2}^{\gamma}$, data; the 
other~\cite{epj:c45:633} also considered data from a previous dijet 
photoproduction analysis published by the ZEUS collaboration~\cite{epj:c23:615}. 
It is the purpose of this analysis to test the effectiveness of each 
parameterization at describing HERA photoproduction data. To this end, the 
present analysis was conducted at higher transverse 
energy relative to previous publications. It is expected that at these high 
transverse energies the predictions of next-to-leading-order (NLO) QCD calculations 
should describe the data well, have smaller uncertainties, and allow a more precise 
discrimination between the different parameterizations of the photon PDFs.
The reduction in statistics associated with moving to higher transverse 
energies was in part compensated by the factor of two increase in luminosity, for 
this independent data sample,  
and the extension to higher pseudorapidity\ZcoosysfnBeta \ of the jet compared to the previous 
analysis~\cite{epj:c23:615}.  

Secondly, the present analysis was designed to provide constraints on the 
proton PDFs. Global fits to determine the proton PDFs continue to be a very active 
and important area of research. 
A common feature of these global fits is a large uncertainty in the gluon PDF 
for high values of $x_{p}$, the fractional momentum at which partons inside 
the proton are probed. At such high values ($x_{p} \gtrsim 0.1$), the gluon 
PDF is poorly constrained and so attempts were made for the present 
investigation to measure cross sections which show particular sensitivity to 
these uncertainties. Recently, the ZEUS collaboration included jet data 
into fits for the proton PDFs~\cite{epj:c42:1}. 

Finally, the difference in azimuthal angle of two jets was considered, as in 
previous measurements of charm and prompt photon photoproduction~\cite{npb:729:492,pl:b511:19}. 
In LO QCD, the cross section as a 
function of the azimuthal difference would simply be a delta function located at 
$\pi$\,radians. However, the presence of higher-order effects leads to extra jets in 
the final state and in values less than 
$\pi$\,radians. The cross section is therefore directly sensitive to higher-order 
topologies and provides a test of NLO QCD and of Monte Carlo (MC) models with 
different implementations of parton-cascade algorithms. The data for charm 
photoproduction~\cite{npb:729:492} demonstrated the inadequacy of NLO QCD, particularly when 
the azimuthal angle difference was significantly different from $\pi$ and for a 
sample of events enriched in resolved-photon processes. To investigate this 
inadequacy in a more inclusive way and with higher precision, such distributions  
were also measured.

\section{Definition of the cross section and variables}
\label{sec-def}

Within the framework of perturbative QCD, the dijet positron-proton cross section, 
$d\sigma_{e p}$, can be written as a convolution of the proton PDFs, 
$f_{p}$, and photon PDFs, $f_{\gamma}$, with the partonic hard cross section, 
$d\hat{\sigma}_{ab}$, as

\begin{equation}
d\sigma_{e p} = \sum_{ab} \int dy f_{\gamma/e} (y)
\int \int dx_{p}dx_{\gamma}f_{p}(x_{p},\mu_F^{2})f_{\gamma}(x_{\gamma},\mu_F^{2})d\hat{\sigma}_{ab}(x_{p},x_{\gamma},\mu_R^{2}), 
\end{equation} 

where $y = E_\gamma/E_e$ is the longitudinal momentum fraction of the almost-real 
photon emitted by 
the positron and the function $f_{\gamma/e}$ is the flux of photons from the positron. 
The equation is a sum over all possible partons, $a$ and $b$.
In the case of the direct cross section, the photon PDF is 
replaced by a delta function at $x_{\gamma}=1$. The scales of the process  
are the renormalization, $\mu_{R}$, and factorization scales, $\mu_{F}$.

To probe the structure of the photon, it is desirable to measure cross sections 
as functions of variables that are sensitive to the incoming parton momentum 
spectrum, such as the momentum fraction, $x_{\gamma}$, at which 
partons inside the photon are probed. Since $x_{\gamma}$ is not directly 
measurable, it is necessary to define~\cite{pl:b322:287} an observable, 
$x_{\gamma}^{\rm obs}$, which is the fraction of the photon momentum participating 
in the production of the two highest transverse-energy jets (and is equal to 
$x_{\gamma}$ for partons in LO QCD), as:

\begin{equation}
x_{\gamma}^{\rm obs} = \frac{E_{T}^{\rm jet1} e^{-\eta^{\rm jet1}} 
                           + E_{T}^{\rm jet2} e^{-\eta^{\rm jet2}}}
                            {2yE_e},
\end{equation}

where $E_e$ is the incident positron energy, $E_{T}^{\rm jet1}$ and 
$E_{T}^{\rm jet2}$ are the transverse energies and $\eta^{\rm jet1}$ and 
$\eta^{\rm jet2}$ the pseudorapidities of the two jets in the laboratory 
frame ($E_{T}^{\rm jet1}>E_{T}^{\rm jet2}$). At LO (see Fig.~\ref{fig:feyn}), direct processes have 
$x_\gamma^{\rm obs} = 1$, while resolved processes have $x_\gamma^{\rm obs} < 1$.

For the proton, the observable $x_p^{\rm obs}$ is similarly defined~\cite{pl:b322:287} 
as

\begin{equation}
x_{p}^{\rm obs} = \frac{E_{T}^{\rm jet1} e^{\eta^{\rm jet1}} 
                      + E_{T}^{\rm jet2} e^{\eta^{\rm jet2}}}{2E_p},
\end{equation}

where $E_{p}$ is the incident proton energy. This observable is the fraction 
of the proton momentum participating in the production of the two 
highest-energy jets (and is equal to $x_{p}$ for partons in LO QCD). 

Cross sections are presented as functions of $x_{\gamma}^{\rm obs}$, 
$x_{p}^{\rm obs}$, $\bar{E}_{T}$, $E_{T}^{\rm jet1}$, $\bar{\eta}$ and $|\Delta\phi^{\rm jj}|$. 
The mean transverse energy of the two jets, $\bar{E}_{T}$, is given by

\begin{equation}
  \bar{E}_{T} = \frac{E_{T}^{\rm jet1} +E_{T}^{\rm jet2}}{2}.
\end{equation}

Similarly, the mean pseudorapidity of the two jets, $\bar{\eta}$, is given by

\begin{equation}
  \bar{\eta} = \frac{\eta^{\rm jet1}+\eta^{\rm jet2}}{2}.
\end{equation}

The absolute difference in azimuthal angle of the two jets, $\phi^{\rm jet1}$ and 
$\phi^{\rm jet2}$, is given by

\begin{equation}
|\Delta\phi^{\rm jj}| = |\phi^{\rm jet1}-\phi^{\rm jet2}|.
\end{equation} 

The kinematic region for this study is defined as 
$Q^{2}<1$\,GeV$^{2}$, where $Q^2 = 2 E_e E_e^\prime (1 +\cos\theta_e)$ and $E_e^\prime$ and 
$\theta_e$ are the energy and angle, respectively, of the scattered positron. The 
photon-proton center-of-mass energy, $W_{\gamma p} = \sqrt{4yE_eE_p}$, 
is required to be in the range $142$ GeV to $293$ GeV. Each event is required to have 
at least two 
jets reconstructed with the $k_{T}$ cluster algorithm~\cite{np:b406:187} in its 
longitudinally invariant inclusive mode~\cite{pr:d48:3160}, with at least one 
jet having transverse energy greater than $20$\,GeV and another greater than 
$15$\,GeV. The jets are required to satisfy \mbox{$-1 < \eta^{\rm jet1,2} < 3$} 
with at least one jet lying in the range between $-1$ and $2.5$. 
The upper bound of 3~units represents an extension of the pseudorapidity range by 
$0.6$ units in the forward direction over the previous analysis~\cite{epj:c23:615}, 
thereby increasing the sensitivity of the measurement to low-$x_\gamma$ and 
high-$x_{p}$ processes. The cross sections for all distributions have been 
determined for regions enriched in direct- and resolved-photon processes by 
requiring \xgo \ to be greater than $0.75$ or less than $0.75$, respectively.

One of the goals of the present investigation is to provide data that constrain 
the gluon PDF in the proton, which exhibits large uncertainties at values of 
$x_{p} \gtrsim 0.1$. A study was performed~\cite{thesis:targett-adams:2006} by 
considering the 
$x_p^{\rm obs}$ cross section in different kinematic regions, varying the cuts 
on the jet transverse energies and pseudorapidities as well as on \xgo. This 
allowed the determination of kinematic regions in which the cross section was 
large enough to be measured and in which the uncertainties on the cross section 
that arise due to those of the gluon PDF were largest.
These cross sections will be referred to as ``optimized'' cross 
sections and are those which have the largest uncertainty from the gluon PDF; 
in total eight cross sections were measured (four direct enriched 
and four resolved enriched). The PDF sets chosen to conduct the optimization 
study were the ZEUS-S~\cite{pr:d67:012007} and ZEUS-JETS~\cite{epj:c42:1} PDF 
sets. The kinematic regions of the cross sections are defined in 
Table~\ref{table:opt}, where the $W_{\gamma p}$ and $Q^2$ requirements are as 
above.

\section{Experimental conditions}
\label{sec-exp}

The data were collected during the 1998-2000 running periods, where 
HERA operated with protons of energy $E_{p} = 920$ GeV and electrons or positrons 
of energy $E_{e} = 27.5$ GeV. During 1998 and the first half of 1999, a sample of 
electron data corresponding to an integrated luminosity of 
$16.7 \pm 0.3$ pb$^{-1}$ was collected. The remaining data up to the year 2000 
were taken using positrons and correspond to an integrated luminosity of 
$65.1 \pm 1.5$ pb$^{-1}$. The results presented here are therefore based on a 
total integrated luminosity of $81.8 \pm 1.8$ pb$^{-1}$. A detailed description 
of the ZEUS detector can be 
found elsewhere~\cite{pl:b297:404,zeus:1993:bluebook}. A brief outline of the 
components that are most relevant for this analysis is given below.

\Zctddesc\ 

\Zcaldesc

The luminosity was measured from the rate of the bremsstrahlung process 
$ep~\rightarrow~e\gamma p$, where the photon was measured in a 
lead--scintillator calorimeter~\cite{desy-92-066,*zfp:c63:391,*acpp:b32:2025} 
placed in the HERA tunnel at $Z=-107~{\rm m}$.

\section{Monte Carlo models}
\label{sec-mc}

The acceptance and the effects of detector response were determined using samples 
of simulated events. The programs 
{\sc Herwig~6.505}~\cite{jhep:0101:010,*cpc:67:465} and 
{\sc Pythia~6.221}~\cite{cpc:135:238,*cpc:82:74}, which implement the 
leading-order matrix 
elements, followed by parton showers and hadronization, were used. The {\sc Herwig} 
and {\sc Pythia} generators differ in the details of the implementation 
of the leading-logarithmic parton-shower models and hence are also compared to 
the measured cross-section $d\sigma/d|\Delta \phi^{\rm jj}|$. The MC programs also 
use different hadronization models:  
{\sc Herwig} uses the cluster model~\cite{np:b238:492} and {\sc Pythia} uses the 
Lund string model~\cite{prep:97:31}. Direct and resolved events were generated 
separately. For the purposes of correction, the relative contribution of direct and resolved 
events was fitted to the data. 
For all generated events, the ZEUS detector response was simulated in 
detail using a program based on {\sc Geant}~3.13~\cite{tech:cern-dd-ee-84-1}.

For both MC programs, the CTEQ5L~\cite{pr:d55:1280} and GRV-LO~\cite{pr:d45:3986,*pr:d46:1973} 
proton and photon PDFs, 
respectively, were used. The $p_T^{\rm min}$ for the outgoing partons from the 
hard scatter was set to 4\,GeV. For the generation of resolved photon events, 
the default multiparton interaction models~\cite{zfp:c72:636,pr:d36:2019} were used. 
A comparably 
reasonable description of the raw data kinematic distributions was observed with 
both {\sc Herwig} and {\sc Pythia} MC simulations.

\section{NLO QCD calculations}
\label{sec-nlo}

The calculation for jet photoproduction used is that of Frixione and 
Ridolfi~\cite{np:b467:399,*np:b507:295,np:b507:315}, which employs the subtraction 
method~\cite{np:b178:421} for dealing with the collinear and infra-red divergencies. 
The number of flavors was set to 5 and the renormalization and factorization 
scales were both set to $\langle E_T^{\rm parton} \rangle$, which is half the sum of 
the transverse energies of the final-state partons. The parton densities in the proton 
were parameterized using CTEQ5M1~\cite{pr:d55:1280}; the value \mbox{$\alpha_s (M_Z) = 0.118$} 
used therein was adopted for the central prediction. 

The following parameterizations of the photon PDFs were used: Cornet et al. 
(CJK)~\cite{pr:d70:093004}, Aurenche et al. (AFG04)~\cite{epj:c34:395}, Slominski et al. 
(SAL)~\cite{epj:c45:633}, Gl\"{u}ck et al. (GRV-HO)~\cite{pr:d45:3986,*pr:d46:1973} 
and a previous set of PDFs from Aurenche et al. (AFG)~\cite{zfp:c64:621}. The three new 
PDFs~\cite{pr:d70:093004,epj:c34:395,epj:c45:633} use all available data on $F_2^\gamma$ from the LEP experiments. The data are of higher precision and cover a wider region of phase space, reaching 
lower in $x_\gamma$ and higher in the momentum of the exchanged photon, compared to the data used in the AFG and 
GRV-HO parameterizations. The parameterization from CJK uses a more careful 
treatment of heavy quarks, whereas that from SAL also considers previous 
dijet photoproduction data from ZEUS~\cite{epj:c23:615}. The most striking 
difference between the resulting PDFs is that CJK has a more rapid rise of 
the gluon density at low $x_\gamma$.

The NLO QCD predictions were corrected for hadronization effects using a 
bin-by-bin procedure according to 
$d\sigma = d\sigma^{\rm NLO} \cdot C_{\rm had}$, where $d\sigma^{\rm NLO}$ is 
the cross section for partons in the final state of the NLO calculation. The 
hadronization correction factor, $C_{\rm had}$, was defined as the 
ratio of the dijet cross sections after and before the hadronization process, 
$C_{\rm had}= d\sigma^{\rm Hadrons}_{\rm MC}/d\sigma^{\rm Partons}_{\rm MC}$. The 
value of $C_{\rm had}$ was taken as the mean of the ratios obtained using the 
{\sc Herwig} and {\sc Pythia} predictions. The hadronization correction was 
generally below $10\%$ in each bin.

Several sources of theoretical uncertainty were investigated, which are given below 
with their typical size,

\begin{itemize}

\item the renormalization scale was changed to 
      ${2}^{\pm 0.5} \cdot \langle E_T^{\rm parton} \rangle$~\cite{epj:c42:1}. This led to an uncertainty of 
      $\mp (10 - 20)\%$;

\item the factorization scale was changed to 
      ${2}^{\pm 0.5} \cdot \langle E_T^{\rm parton} \rangle$~\cite{epj:c42:1}. This led to an uncertainty of 
      $\pm (5 - 10)\%$;

\item the value of $\alpha_s$ was changed by $\pm 0.001$, the uncertainty on the world 
      average~\cite{ppnp:58:351}, by using the CTEQ4 PDFs for $\alpha_s(M_Z) =$ 0.113,  
      0.116 and 0.119 and interpolating accordingly. This led to an uncertainty 
      of about $\pm 2\%$;

\item the uncertainty in the hadronization correction was estimated as half the 
      spread between the two MC correction factors. This led to an uncertainty of 
      generally less than $\pm 5\%$.

\end{itemize}

\begin{figure}[htb]
\begin{center}
~\epsfig{file=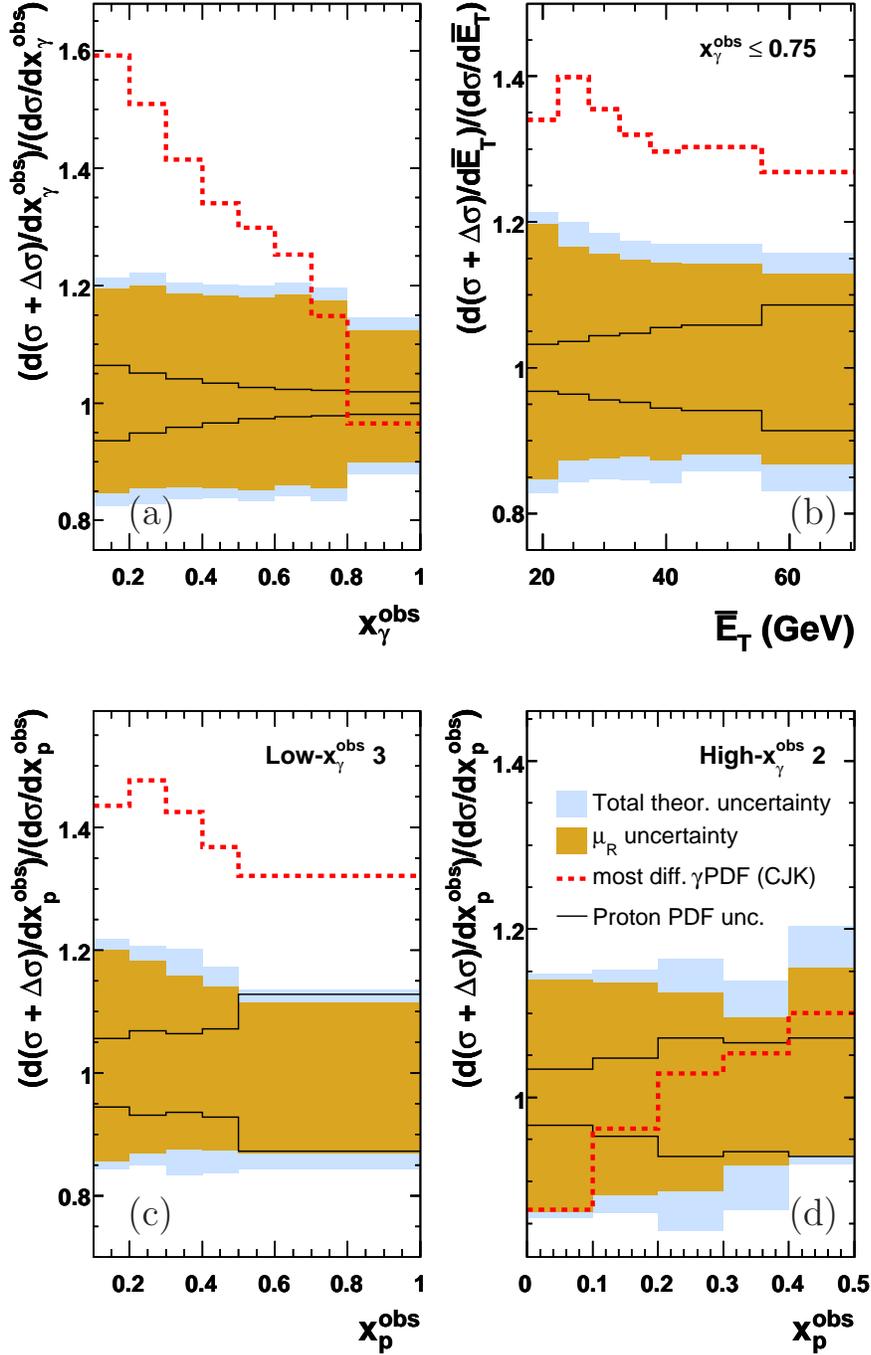,height=18cm}
\put(-280,325){\makebox(0,0)[tl]{\large (a)}}
\put(-30,325){\makebox(0,0)[tl]{\large (b)}}
\put(-280,60){\makebox(0,0)[tl]{\large (c)}}
\put(-30,60){\makebox(0,0)[tl]{\large (d)}}
\caption{The theoretical uncertainties (see Section~\ref{sec-nlo}) for sample distributions: 
(a) \xgo, (b) $\bar{E}_T$ for \xgo $\leq$ 0.75, (c) ``Low-\xgo \ 3'' and (d) ``High-\xgo \ 2'', 
which are defined in Table~\ref{table:opt}. The uncertainties are the total (outer shaded 
band), that from varying $\mu_R$ (inner shaded band), the experimental uncertainties of data 
propagated in the ZEUS-JETS fit (solid lines) and using the most different photon PDF, CJK 
(dashed line) instead of AFG04.}
\label{fig:theo-unc}
\end{center}
\end{figure}

The above four uncertainties were added in quadrature and are displayed on the figures 
as the shaded band around the central prediction. The size of these uncertainties is 
also shown as a function of $\bar{E}_T$, \xgo \ and $x_p^{\rm obs}$ in 
Fig.~\ref{fig:theo-unc}. The uncertainty from changing the renormalization scale is 
dominant. It should be noted that here the renormalization and factorization scales 
were varied independently by factors of ${2}^{\pm 0.5}$ and the resulting changes 
were added in quadrature as in the determination of the ZEUS-JETS 
PDF~\cite{epj:c42:1}. The result of this procedure leads to an uncertainty which is 
approximately the same as varying both simultaneously by ${2}^{\pm 1}$ as has been 
done previously~\cite{epj:c23:615}.

Other uncertainties which were considered are:

\begin{itemize}

\item the uncertainties in determining the proton PDFs were assessed by using the 
      ZEUS-JETS PDF uncertainties propagated from the experimental uncertainties of 
      the fitted data. This led to an uncertainty of $\pm (5-10) \%$;

\item the uncertainties in determining the photon PDFs were assessed by using 
      sets from different authors. Differences of generally less than 25\% were observed 
      between the AFG, AFG04, SAL and GRV sets. However, the predictions based on CJK  
      were up to 70\% higher than those based on the other four.

\end{itemize}

\begin{figure}[htb]
\begin{center}
~\epsfig{file=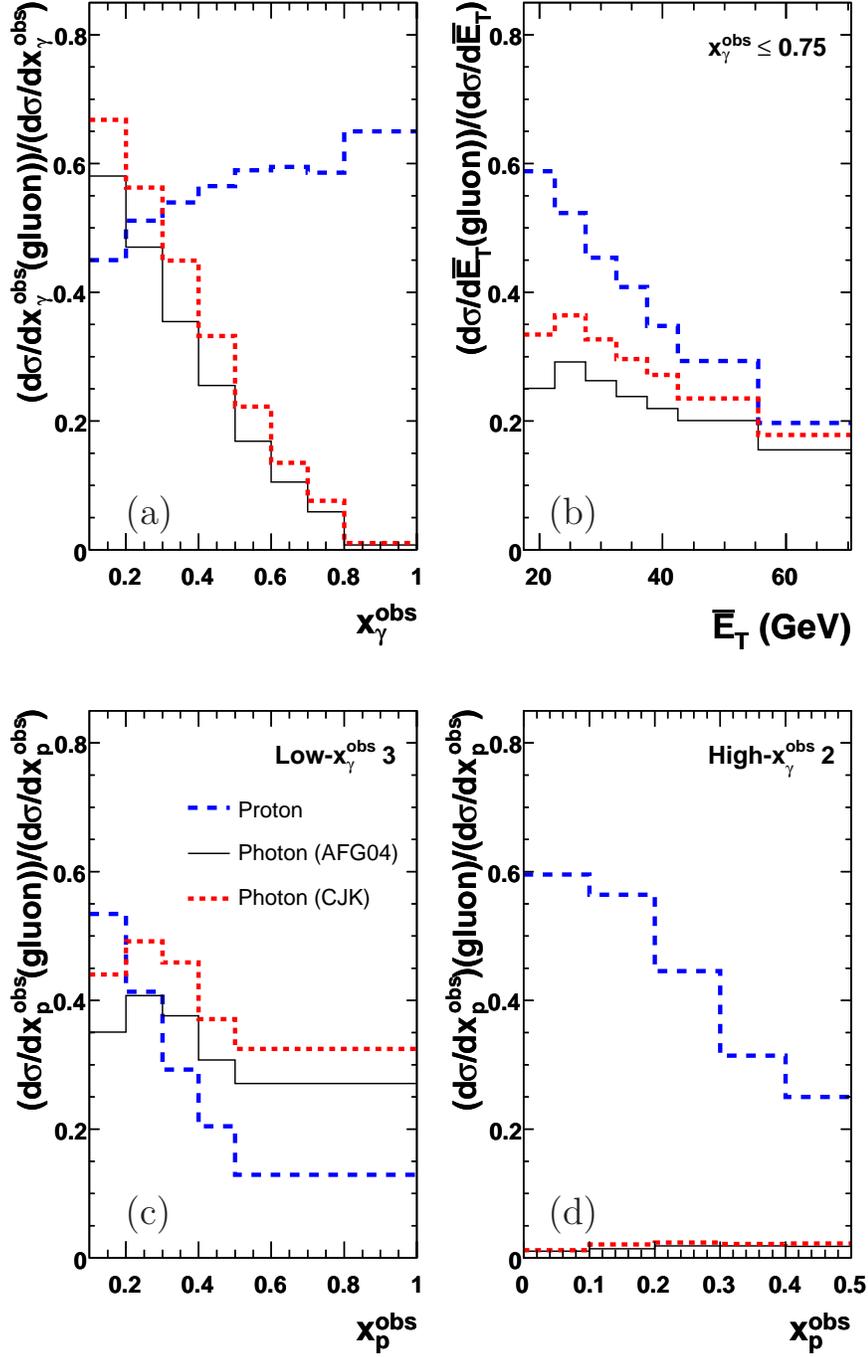,height=18cm}
\put(-280,325){\makebox(0,0)[tl]{\large (a)}}
\put(-120,325){\makebox(0,0)[tl]{\large (b)}}
\put(-280,60){\makebox(0,0)[tl]{\large (c)}}
\put(-120,60){\makebox(0,0)[tl]{\large (d)}}
\caption{Predictions of the fraction of the cross section initiated by gluons for sample 
distributions: (a) \xgo, (b) $\bar{E_T}$ for \xgo $\leq$ 0.75, (c) ``Low-\xgo \ 3'' and 
(d) ``High-\xgo \ 2'', which are defined in Table~\ref{table:opt}. The gluon fractions are 
from the proton using the CTEQ5M1 PDF (long-dashed line), and from the photon using the 
AFG04 (solid line) and CJK PDFs (short-dashed line).}
\label{fig:gluon}
\end{center}
\end{figure}

These uncertainties were not added in quadrature with the others, but examples of their size are given  
in Fig.~\ref{fig:theo-unc}. Differences between the two photon PDFs, CJK and AFG04, 
are concentrated at low \xgo \ and low $\bar{E}_T$; the low \xgo \ region is most 
sensitive to the gluon distribution in the photon, which increases more rapidly for 
CJK as shown in Fig.~\ref{fig:gluon}. At lowest \xgo, the fraction of the cross section 
arising from the gluon distribution in the photon is 66\% for CJK. The uncertainty on 
the proton PDF increases with increasing $\bar{E}_T$ and $x_p^{\rm obs}$ and is sometimes, 
as seen in Fig.~\ref{fig:theo-unc}(c), as large as the other combined uncertainties. 
The fraction of the cross section arising from the gluon distribution in the proton 
is about 50\% for the lower $\bar{E}_T$ and $x_p^{\rm obs}$ values considered, 
but decreases to below 20\% for high values. However, the uncertainty on the gluon 
dominates the proton PDF uncertainty in most of the kinematic region investigated.

\section{Event selection}
\label{sec-event}

A three-level trigger system was used to select events 
online~\cite{zeus:1993:bluebook,uproc:chep:1992:222,epj:c1:109}. At the third 
level, a cone algorithm was applied to the CAL cells and jets were reconstructed 
using the energies and positions of these cells. Events with at least one jet, 
which satisfied the requirements that the transverse energy exceeded $10$ GeV 
and the pseudorapidity was less than $2.5$, were accepted. Dijet events in 
photoproduction were then selected offline by using the following procedures 
and cuts designed to remove sources of background:

\begin{itemize}

\item to remove background due to proton beam-gas interactions and cosmic-ray 
      showers, the longitudinal position of the reconstructed vertex was 
      required to be in the range $|Z_{\rm vertex}|<40$\,cm;

\item a cut on the ratio of the number of tracks assigned to the primary vertex 
      to the total number of tracks, $N_{\rm trk}^{\rm vtx}/N_{\rm trk}>0.1$, 
      was also imposed to remove beam-related background, which have values of 
      this ratio typically below 0.1;

\item to remove background due to charged current deep inelastic scattering (DIS) 
      and cosmic-ray showers, events were required to have a relative 
      transverse momentum of 
      $p_T/\sqrt{E_T} < 1.5\,\sqrt{\rm GeV}$, where $p_T$ and $E_T$ are, 
      respectively, the measured transverse momentum and transverse energy of the 
      event;

\item neutral current (NC) DIS events with a scattered positron candidate in the CAL 
      were removed by cutting~\cite{pl:b322:287} on the inelasticity, $y$, which 
      is estimated from the energy, $E_e^\prime$, and polar angle, $\theta_e^\prime$, 
      of the scattered positron candidate using 
      \mbox{$y_e=1-\frac{E_e^\prime}{2E_e}(1-\cos\theta_e^\prime)$}. Events were rejected if 
      $y_e < 0.7$;

\item the requirement $0.15<y_{\rm JB}<0.7$ was imposed, where $y_{\rm JB}$ is 
      the estimator of $y$ measured from the CAL energy deposits according to the 
      Jacquet-Blondel method~\cite{proc:epfacility:1979:391}. The upper cut 
      removed NC DIS events where the positron was not identified and which therefore 
      have a value of $y_{\rm JB}$ close to 1. The lower cut removed proton 
      beam-gas events which typically have a low value of $y_{\rm JB}$; 

\item the $k_T$-clustering algorithm was applied to the CAL energy deposits. 
      The transverse energies of the jets were 
      corrected~\cite{epj:c23:615,pl:b531:9,proc:calor02:2002:767} in order to compensate 
      for energy losses in inactive material in front of the CAL. Events 
      were selected in which at least two jets were found with 
      $E_{T}^{\rm jet1} >$~20\,GeV, $E_{T}^{\rm jet2} >$~15\,GeV and 
      $-1<\eta^{\rm jet1,2}<3$, with at least one jet lying in the range between 
      $-1$ and $2.5$. In this region, the resolution of the jet transverse energy was 
      about 10\%.

\end{itemize}

\section{Data correction and systematics}
\label{sec-unc}

The data were corrected using the MC samples detailed in Section~\ref{sec-mc} 
for acceptance and the effects of detector response using the bin-by-bin method, 
in which the correction factor, as a function of an observable ${\mathcal O}$ in a 
given bin $i$, is 
$C_i({\mathcal O})=N_i^{\rm had}({\mathcal O})/N_i^{\rm det}({\mathcal O})$. 
The variable $N_i^{\rm had}({\mathcal O})$ is the number of events in the simulation 
passing the kinematic requirements on the hadronic final state described in 
Section~\ref{sec-def} and $N_i^{\rm det}({\mathcal O})$ is the number of 
reconstructed events passing the selection requirements as detailed in 
Section~\ref{sec-event}. 

The results of a detailed analysis~\cite{thesis:targett-adams:2006,thesis:perrey:2007} 
of the possible sources of systematic uncertainty are listed below. Typical values for 
the systematic uncertainty are quoted for the cross sections as a function of \xgo, 

\begin{itemize}

\item varying the measured jet energies by $\pm1\%$~\cite{epj:c23:615,pl:b531:9,proc:calor02:2002:767} in the simulation, 
      in accordance with the uncertainty in the jet energy scale, gave an 
      uncertainty of $\mp 5\%$;

\item the central correction factors were determined using the {\sc Pythia} MC. 
      The {\sc Herwig} MC sample was used to assess the model dependency of this 
      correction and gave an uncertainty of +4\%, but up to +12\% at lowest 
      $x_\gamma^{\rm obs}$;

\item changing the values of the various cuts to remove backgrounds from DIS, 
      cosmic-ray and beam-gas events gave a combined uncertainty of less than 
      $\pm 1\%$;

\item varying the fraction of direct processes between 34\% and 70\% of the total MC 
      sample in order to describe each of the kinematic distributions gave an 
      uncertainty of about $^{+2}_{-5}\%$;

\item changing the proton and photon PDFs to CTEQ4L~\cite{pr:d55:1280} and 
      WHIT2~\cite{pr:d51:3197} respectively in the MC samples gave an uncertainty of 
      about $\pm 1.5\%$ and $\pm 2.5\%$.

\end{itemize}

\begin{figure}[htb]
\begin{center}
~\epsfig{file=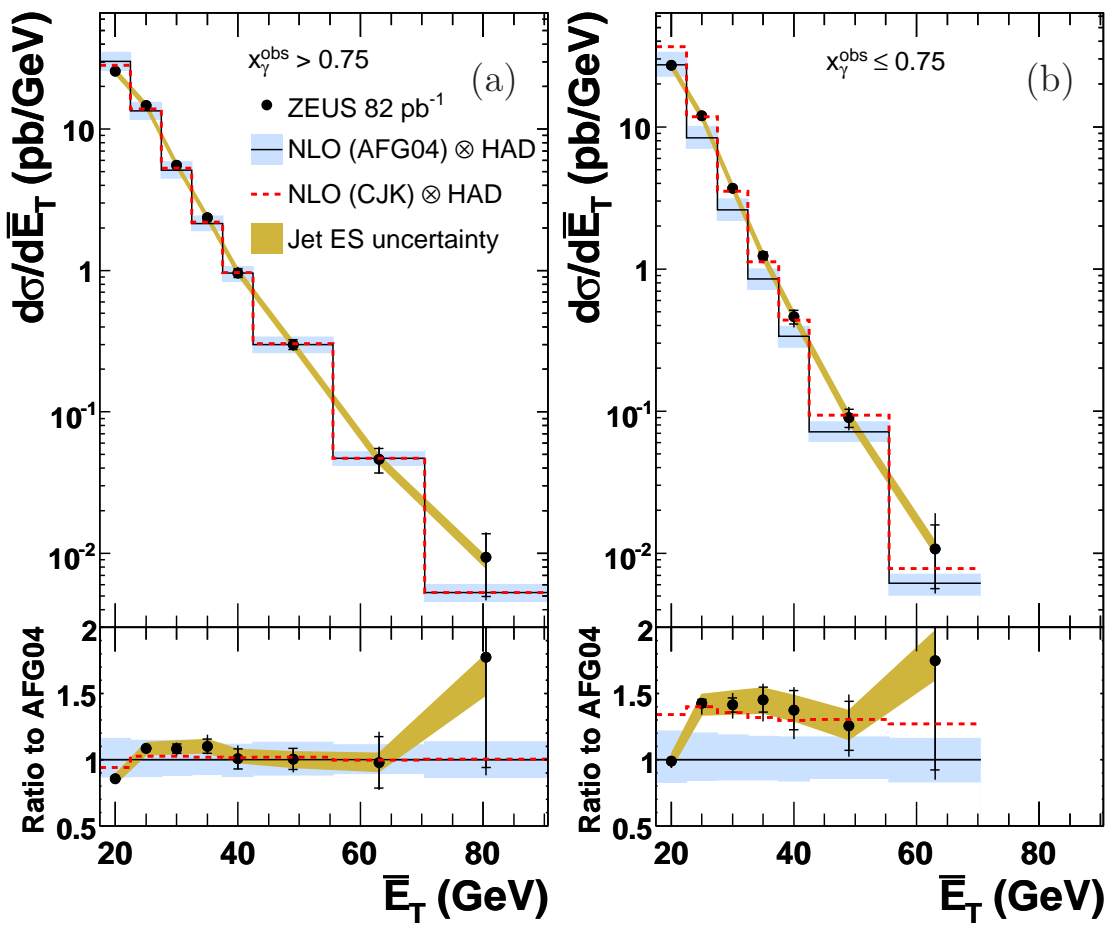,height=14.5cm}
\put(-245,345){\makebox(0,0)[tl]{\large (a)}}
\put(-35,345){\makebox(0,0)[tl]{\large (b)}}
\caption{Measured cross-section $d\sigma/d\bar{E_T}$ for (a) \xgo $>$ 0.75 and 
(b) \xgo $\leq$ 0.75 compared with NLO QCD predictions using the AFG04 (solid line) 
and CJK (dashed line) photon PDFs. The data (dots) are shown with statistical (inner 
bars) and statistical and systematic uncertainties added in quadrature (outer bars) 
along with the jet energy-scale (Jet ES) uncertainty (shaded band). The NLO QCD 
predictions are shown (NLO QCD $\otimes$ HAD) multiplied by the hadronization 
corrections, $C_{\rm had}$, discussed in Section~\ref{sec-nlo}. The predictions using 
AFG04 are 
also shown with their associated uncertainties (shaded histogram) as discussed in 
Section~\ref{sec-nlo}. The ratios to the prediction using the AFG04 photon PDF are 
shown at the bottom of the figure.}
\label{fig:gen_et}
\end{center}
\end{figure}

\begin{figure}[htb]
\begin{center}
~\epsfig{file=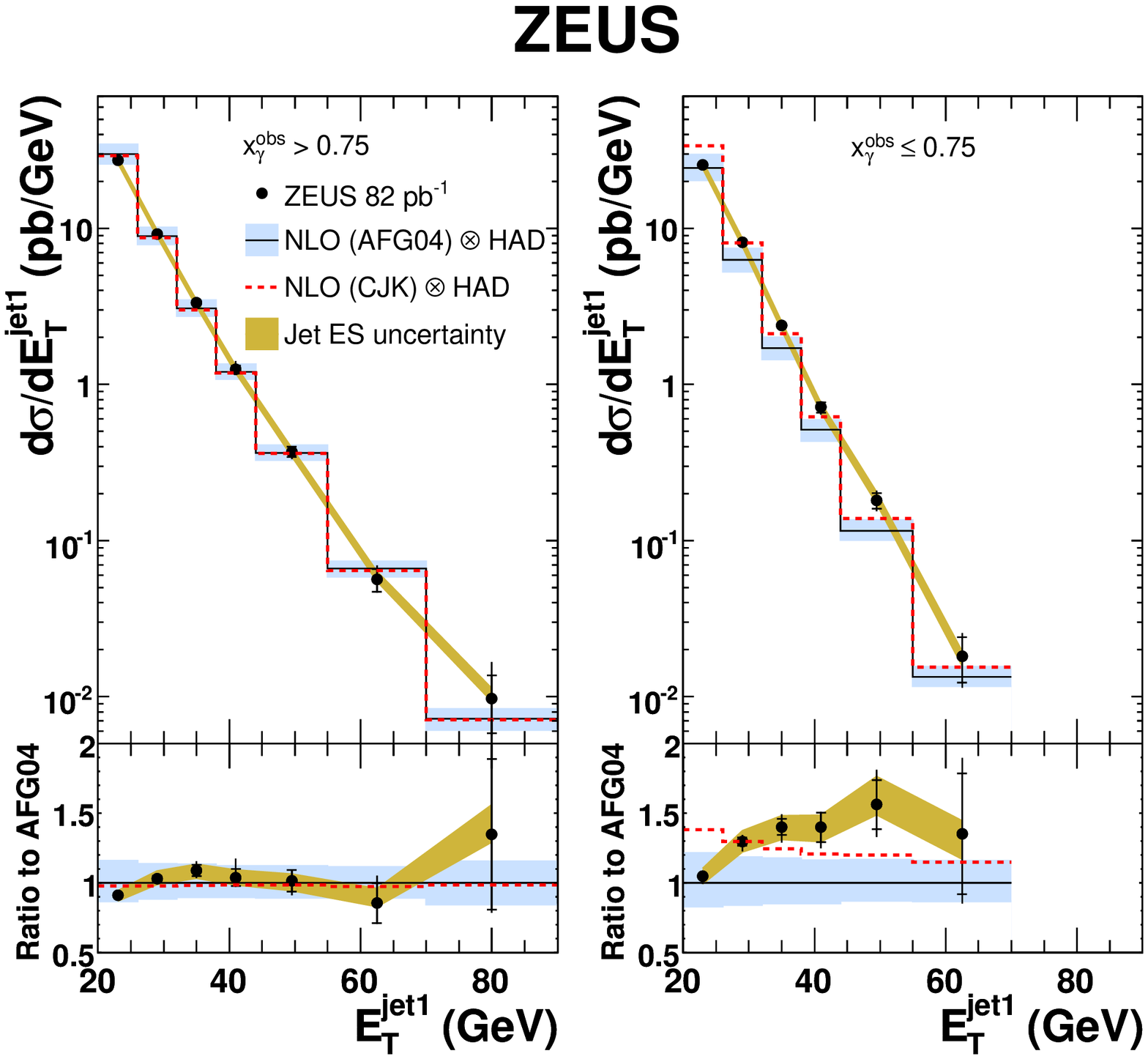,height=14.5cm}
\put(-245,345){\makebox(0,0)[tl]{\large (a)}}
\put(-35,345){\makebox(0,0)[tl]{\large (b)}}
\caption{Measured cross-section $d\sigma/dE_T^{\rm jet1}$ for (a) \xgo $>$ 0.75 and (b) 
\xgo $\leq$ 0.75. For further details, see the caption to Fig.~\ref{fig:gen_et}.}
\label{fig:gen_etjet1}
\end{center}
\end{figure}

\begin{figure}[htb]
\begin{center}
~\epsfig{file=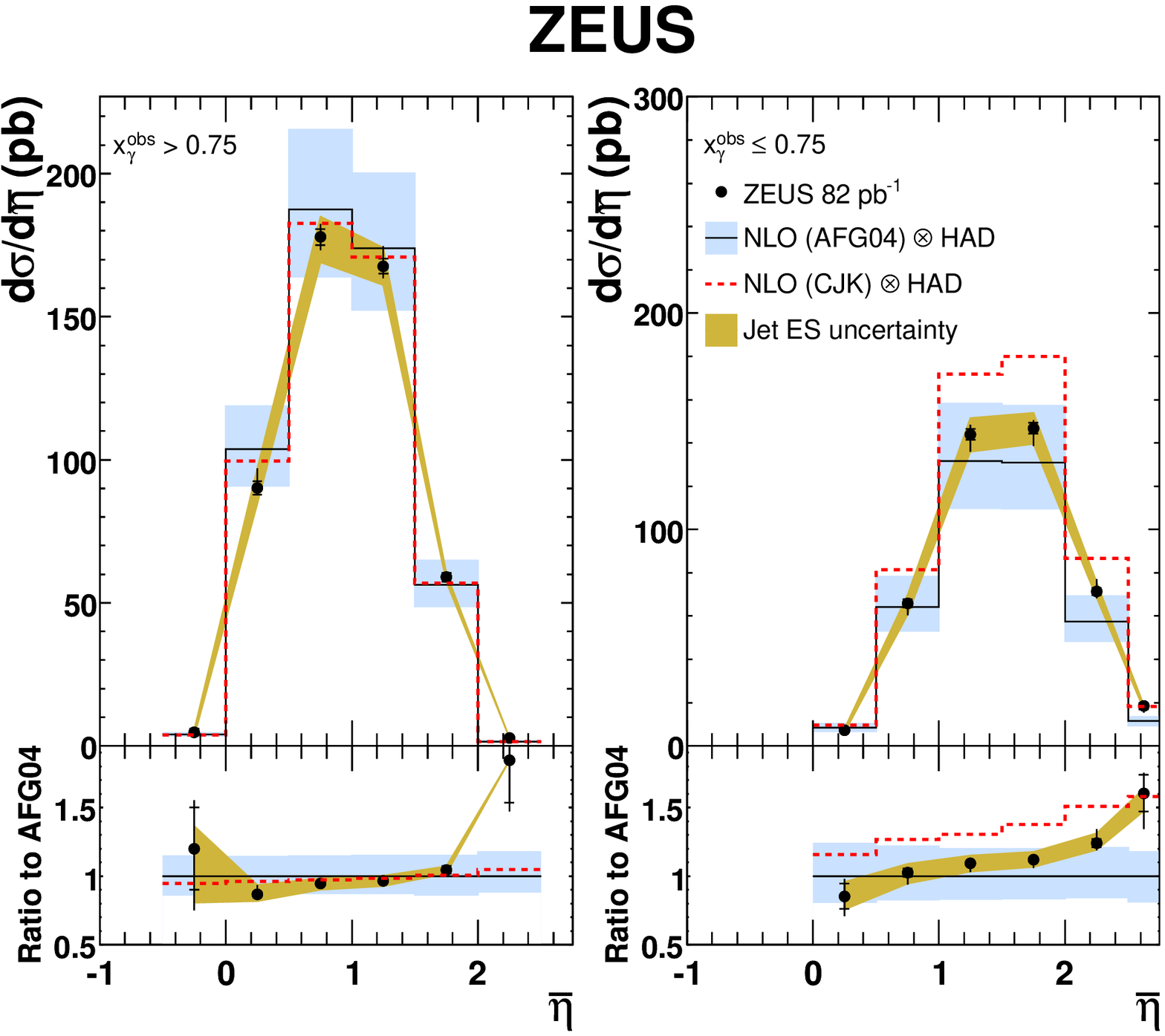,height=14.5cm}
\put(-245,345){\makebox(0,0)[tl]{\large (a)}}
\put(-35,345){\makebox(0,0)[tl]{\large (b)}}
\caption{Measured cross-section $d\sigma/d\bar{\eta}$ for (a) \xgo $>$ 0.75 and (b) 
\xgo $\leq$ 0.75. For further details, see the caption to Fig.~\ref{fig:gen_et}.}
\label{fig:gen_eta}
\end{center}
\end{figure}

\begin{figure}[htb]
\begin{center}
~\epsfig{file=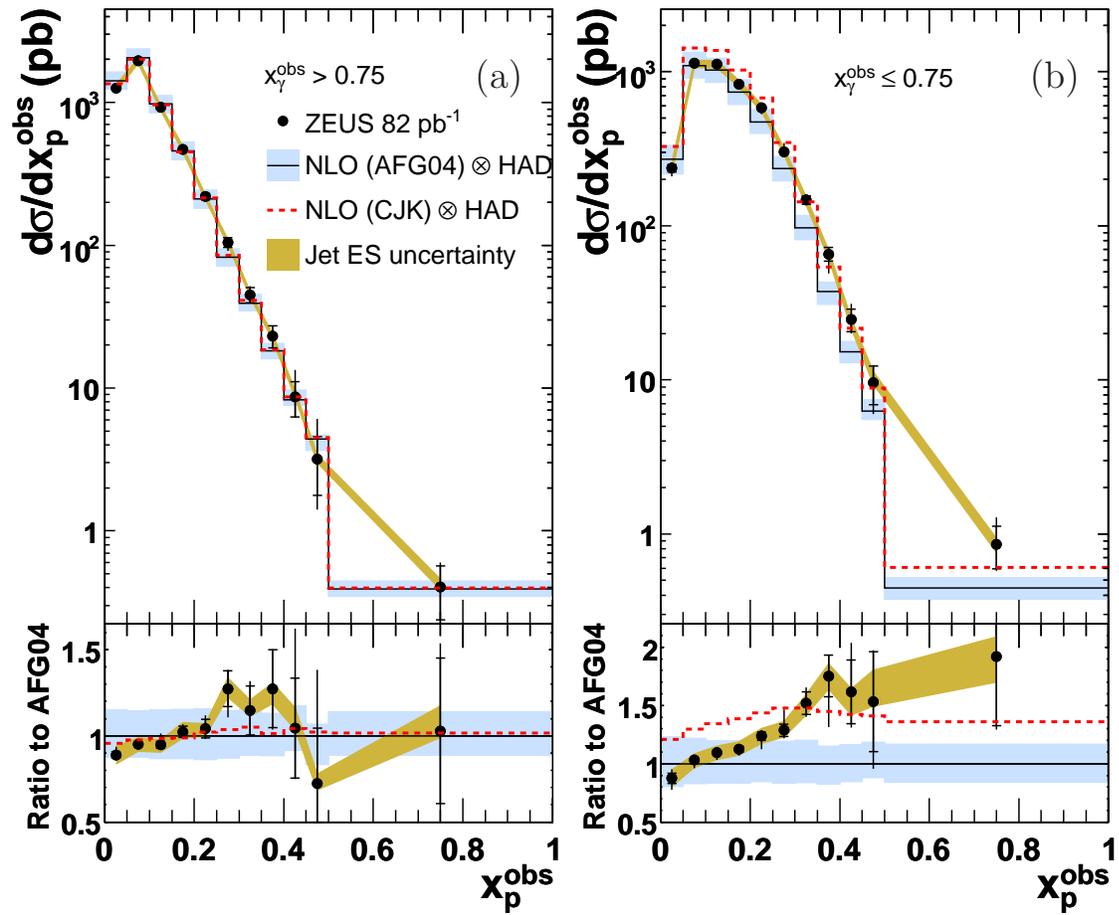,height=14.5cm}
\put(-245,345){\makebox(0,0)[tl]{\large (a)}}
\put(-35,345){\makebox(0,0)[tl]{\large (b)}}
\caption{Measured cross-section $d\sigma/dx_p^{\rm obs}$ for (a) \xgo $>$ 0.75 and (b) 
\xgo $\leq$ 0.75. For further details, see the caption to Fig.~\ref{fig:gen_et}.}
\label{fig:gen_xp}
\end{center}
\end{figure}

\begin{figure}[htb]
\begin{center}
~\epsfig{file=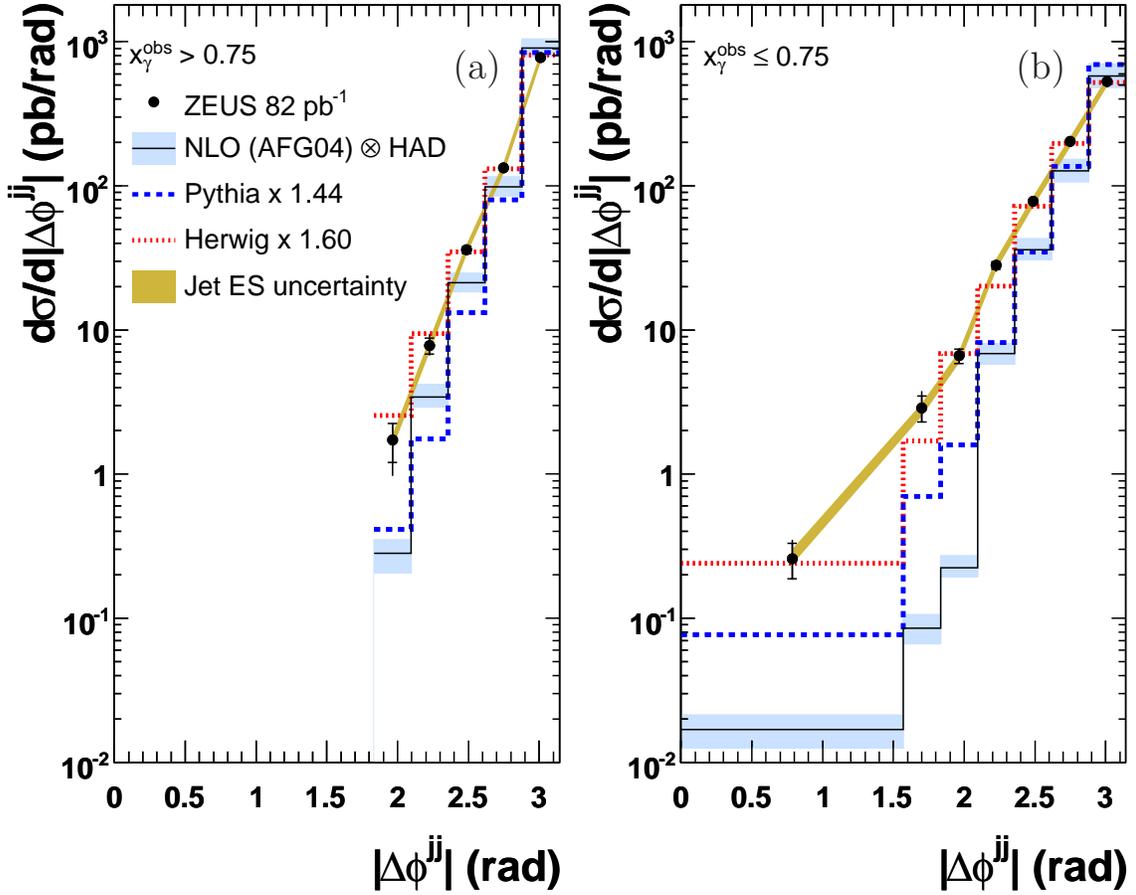,height=13.5cm}
\put(-260,322){\makebox(0,0)[tl]{\large (a)}}
\put(-47,322){\makebox(0,0)[tl]{\large (b)}}
\caption{Measured cross-section $d\sigma/d\Delta |\phi^{\rm jj}|$ for (a) \xgo $>$ 0.75 and 
(b) \xgo $\leq$ 0.75 compared with NLO QCD predictions using the AFG04 (solid line) photon 
PDF. Predictions from the MC programs {\sc Herwig} (dot-dashed) and {\sc Pythia} 
(dashed), area normalized to the data by the factors given, are also shown. The data (dots) 
are shown with statistical (inner bars) and 
statistical and systematic uncertainties added in quadrature (outer bars) along with the 
jet energy-scale (Jet ES) uncertainty (shaded band). The NLO QCD 
predictions are shown (NLO QCD $\otimes$ HAD) multiplied by the hadronization 
corrections, $C_{\rm had}$, discussed in Section~\ref{sec-nlo}. The predictions using 
AFG04 are also shown 
with their associated uncertainties (shaded histogram) as discussed in Section~\ref{sec-nlo}.}
\label{fig:gen_phi}
\end{center}
\end{figure}

\begin{figure}[htb]
\begin{center}
~\epsfig{file=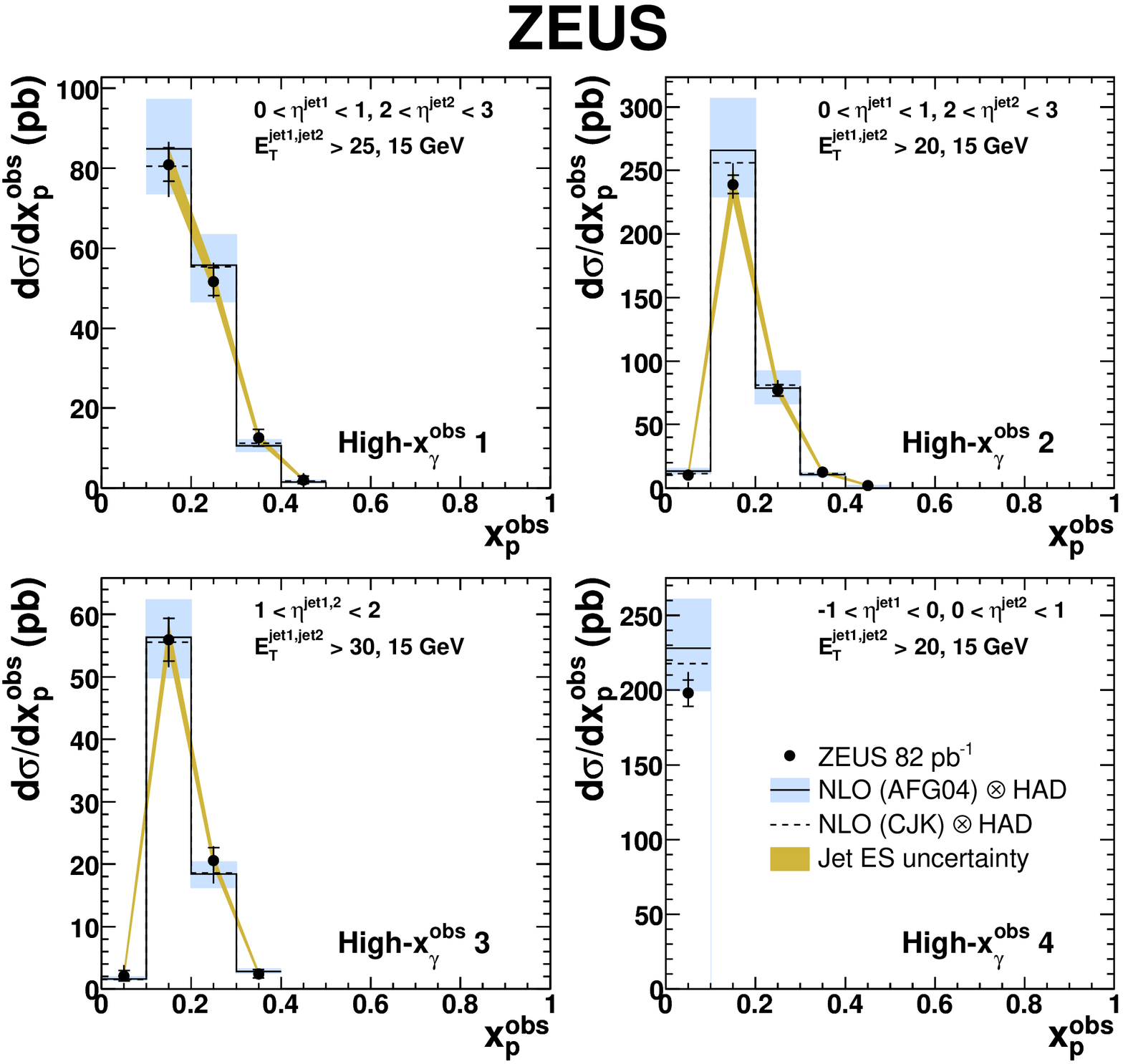,height=14.5cm}
\caption{Optimized cross-sections $d\sigma/dx_p^{\rm obs}$ for \xgo $>$ 0.75 in the 
kinematic regions defined in Table~\ref{table:opt}. For further details, see the caption 
to Fig.~\ref{fig:gen_et}.}
\label{fig:opt1}
\end{center}
\end{figure}

\begin{figure}[htb]
\begin{center}
~\epsfig{file=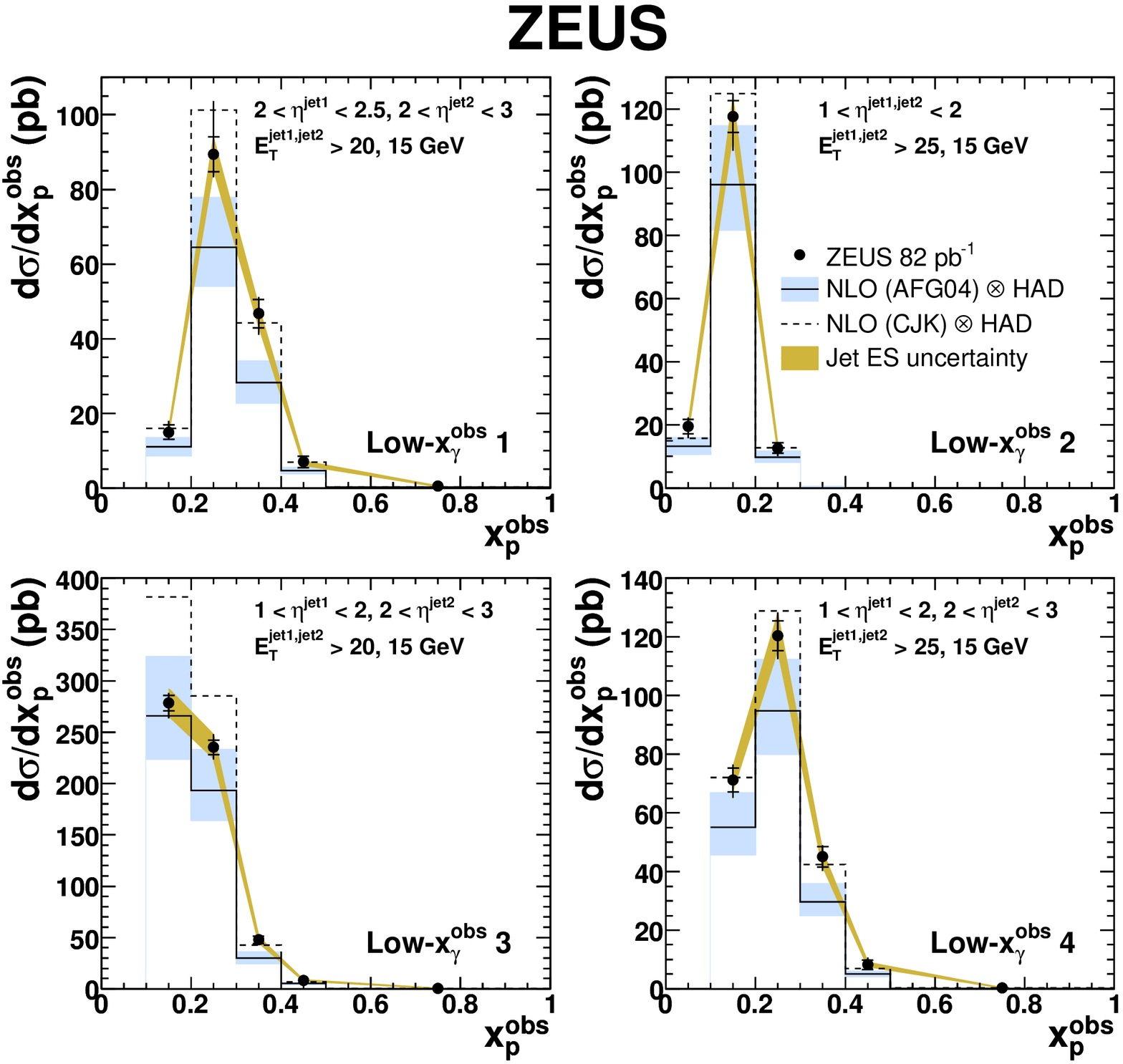,height=14.5cm}
\caption{Optimized cross-sections $d\sigma/dx_p^{\rm obs}$ for \xgo $\leq$ 0.75 in the 
kinematic regions defined in Table~\ref{table:opt}. For further details, see the caption 
to Fig.~\ref{fig:gen_et}.}
\label{fig:opt2}
\end{center}
\end{figure}

The uncertainty in the cross sections due to the jet energy-scale uncertainty is 
correlated between bins and is therefore displayed separately as a shaded band in 
Figs.~\ref{fig:gen_et}--\ref{fig:photon_others}. All other systematic uncertainties
 were added in quadrature when displayed 
in these figures. The choice of MC sample also exhibited some correlation between bins 
and is hence given separately in Tables~\ref{tab:et-high}--\ref{tab:xgamma}. In 
addition, an overall normalization uncertainty of $2.2\%$ from the luminosity 
determination is not included in either the figures or tables.

\section{Results}
\label{sec-res}

\subsection{Dijet differential cross sections}
\label{sec-res1}

Differential cross-sections $d\sigma/d\bar{E}_T$, $d\sigma/dE_T^{\rm jet1}$, $d\sigma/d\bar{\eta}$ and 
$d\sigma/dx_p^{\rm obs} $ are given in Tables~\ref{tab:et-high}--\ref{tab:xp-low} 
and shown in Figs.~\ref{fig:gen_et}--\ref{fig:gen_xp} for \xgo \ above and below 
0.75. For \xgo $>$ 0.75, $d\sigma/d\bar{E}_T$ and $d\sigma/dE_T^{\rm jet1}$ fall by over three orders of 
magnitude over the $\bar{E}_T$ and $E_T^{\rm jet1}$ ranges measured and the jets are produced up 
to $\bar{\eta} \sim$ 2. For \xgo $\leq$ 0.75, the slopes of $d\sigma/d\bar{E}_T$ and $d\sigma/dE_T^{\rm jet1}$ are 
steeper, with the jets produced further forward in $\bar{\eta}$. It is interesting 
to note that in both regions of \xgo, the data probe high values of $x$ in the 
proton.

The NLO QCD predictions, corrected for hadronization and using the AFG04 and CJK photon 
PDFs, are compared to the data. For \xgo $>$ 0.75, the NLO QCD predictions describe 
the data well, although some differences in shape are observed for 
$d\sigma/d\bar{E}_T$ and $d\sigma/dE_T^{\rm jet1}$. Although measurements at 
high \xgo \ are less sensitive to the structure of the photon, it is interesting 
to note that the prediction using the CJK photon PDF describes the $\bar{E}_T$ 
spectrum somewhat better. The shapes for the $\bar{\eta}$ and $x_p^{\rm obs}$ distributions 
are also better reproduced using the CJK photon PDF.

At low \xgo, the difference in shapes between data and NLO QCD for $d\sigma/d\bar{E}_T$ 
and $d\sigma/dE_T^{\rm jet1}$ is more marked, as has been seen previously~\cite{epj:c23:615}. For the prediction 
using AFG04, the data and NLO agree in the lowest bin whereas the prediction is 
significantly lower at higher $\bar{E}_T$ and $E_T^{\rm jet1}$. In contrast, the prediction from CJK is too 
high in the first bin, which dominates the cross section, but agrees well at higher 
$\bar{E}_T$ and $E_T^{\rm jet1}$. For the $\bar{\eta}$ and $x_p^{\rm obs}$ distributions, the shapes are 
again better described by NLO QCD using the CJK photon PDF, although the normalization 
is too high. Sensitivity to the photon PDFs is discussed further in Section~\ref{sec-res4}.

\subsection{Measurement of \boldmath $d\sigma/d|\Delta \phi^{\rm jj}|$}
\label{sec-res2}

The cross-section $d\sigma/d|\Delta \phi^{\rm jj}|$ is presented for \xgo \ above 
and below 0.75 in Tables~\ref{tab:phi-high} and~\ref{tab:phi-low} and 
Fig.~\ref{fig:gen_phi}. For \xgo $>$ 0.75, the cross-section data fall by 
about three orders of magnitude in the cross section, more steeply than for 
\xgo $\leq$ 0.75. The predictions from NLO QCD and also both {\sc Herwig} and {\sc Pythia} 
MC programs (plotted separately since the implementation of parton showers differs 
between the two programs) are compared to the data. The MC predictions are area 
normalized to the data in the measured kinematic region. At high \xgo, NLO QCD agrees with 
the data at highest $|\Delta \phi^{\rm jj}|$, but it has a somewhat steeper fall off. 
The prediction from the {\sc Pythia} MC program is similar to that for NLO QCD, 
whereas the prediction from the {\sc Herwig} program describes the data well. 
For low \xgo, the distribution for NLO QCD is much too steep and is significantly 
below the data for all values of $|\Delta \phi^{\rm jj}|$ except the highest 
bin. The prediction from the {\sc Pythia} program is less steep, but still 
gives a poor description. The prediction from the {\sc Herwig} program is in 
remarkable agreement with the data.

The results and conclusions shown are qualitatively similar to those already 
seen in dijet photoproduction in which at least one of the jets was tagged as 
originating from a charm quark~\cite{npb:729:492}. The results here confirm 
that the parton-shower model in {\sc Herwig} gives a good simulation of 
high-order processes and suggest that a matching of it to NLO QCD would 
give a good description of the data in both shape and normalization. Should such 
a calculation or other high-order prediction become available, the distributions 
presented here would be ideal tests of their validity as they present inclusive 
quantities and also have higher precision compared to the previous  
result~\cite{npb:729:492}.

\subsection{Optimized cross sections}
\label{sec-res3}

The cross-sections $d\sigma/dx_p^{\rm obs}$, optimized to be most sensitive to the 
uncertainty on the gluon PDF in the proton, are given in 
Tables~\ref{tab:optd}--\ref{tab:optg} and shown in Figs.~\ref{fig:opt1} 
and~\ref{fig:opt2} for \xgo \ above and below 0.75, respectively. The measurements 
cover a range in $x_p^{\rm obs}$ of about 0.1 to 0.5. At high \xgo, the data are 
very well described by NLO QCD predictions. At low \xgo, the description by NLO 
QCD is poorer, particularly when using the AFG04 photon PDF. Generally the 
predictions with CJK describe the data better with the exception of the 
\mbox{``Low-$x_\gamma^{\rm obs}$ 3''} cross section. Inclusion of these high-\xgo \ 
data in future fits would constrain the proton PDFs further, in particular that of 
the gluon. To include the cross sections for low \xgo, a systematic treatment 
of the photon PDFs and their uncertainty is needed.

\subsection{Sensitivity to the photon PDFs}
\label{sec-res4}

\begin{figure}[htb]
\begin{center}
\vspace{-1cm}
~\epsfig{file=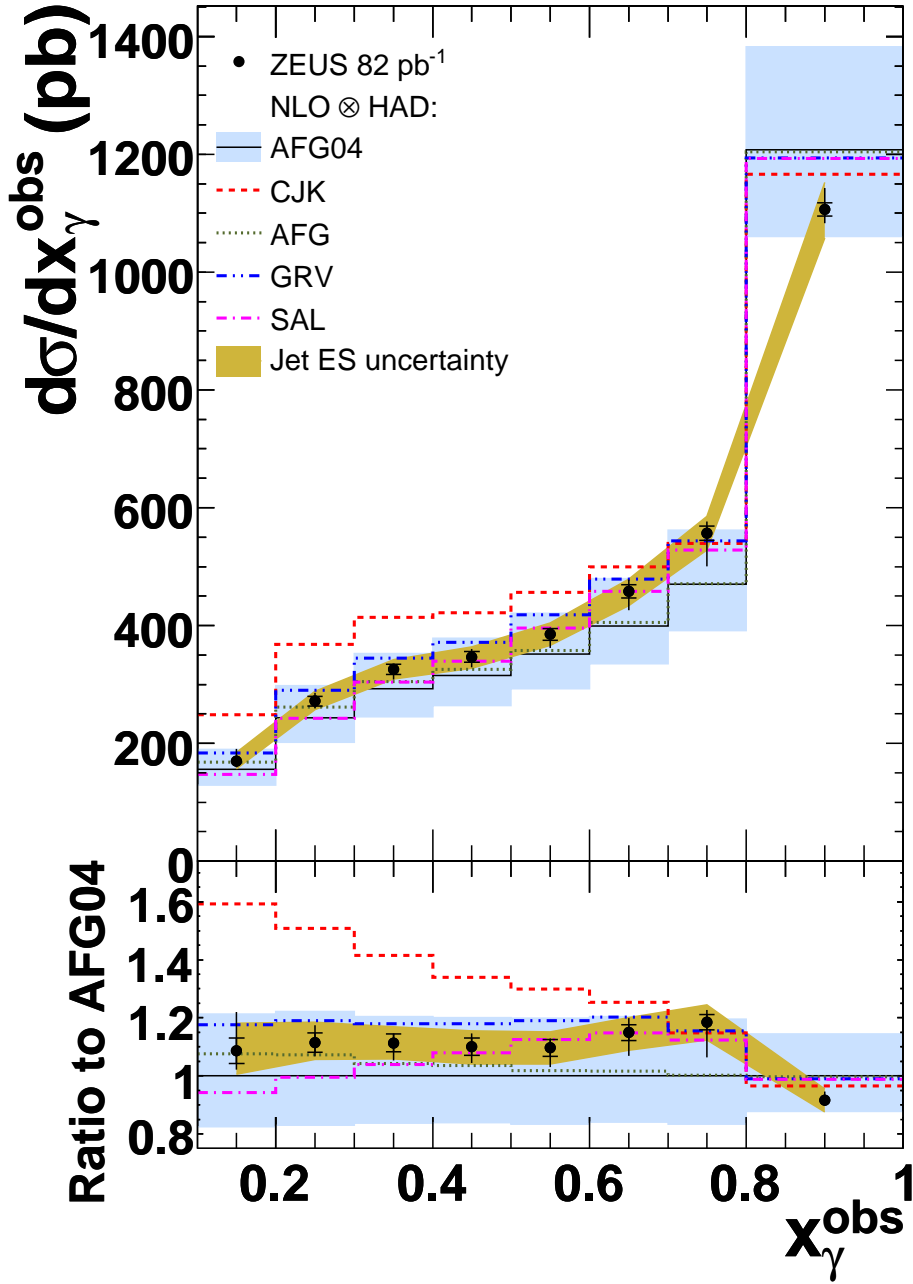,height=20cm}
\vspace{-1cm}
\caption{Measured cross-section for $d\sigma/dx_\gamma^{\rm obs}$ compared with NLO QCD 
predictions using the AFG04 (solid line), CJK (dashed line), AFG (dotted line), 
GRV (dashed and double-dotted line) and SAL (dashed and single-dotted line) photon PDFs. 
The data (dots) are shown with statistical (inner bars) and statistical and 
systematic uncertainties added in quadrature (outer bars) along with the 
jet energy-scale (Jet ES) uncertainty (shaded band). The NLO QCD predictions are 
shown (NLO QCD $\otimes$ HAD) multiplied by the hadronization corrections, 
$C_{\rm had}$, discussed in Section~\ref{sec-nlo}. The predictions using AFG04 are 
also shown with their associated uncertainties (shaded histogram) as discussed in 
Section~\ref{sec-nlo}. The ratios to the prediction using the AFG04 photon PDF are 
shown at the bottom of the figure.}
\label{fig:photon_xgamma}
\end{center}
\end{figure}

\begin{figure}[htb]
\begin{center}
~\epsfig{file=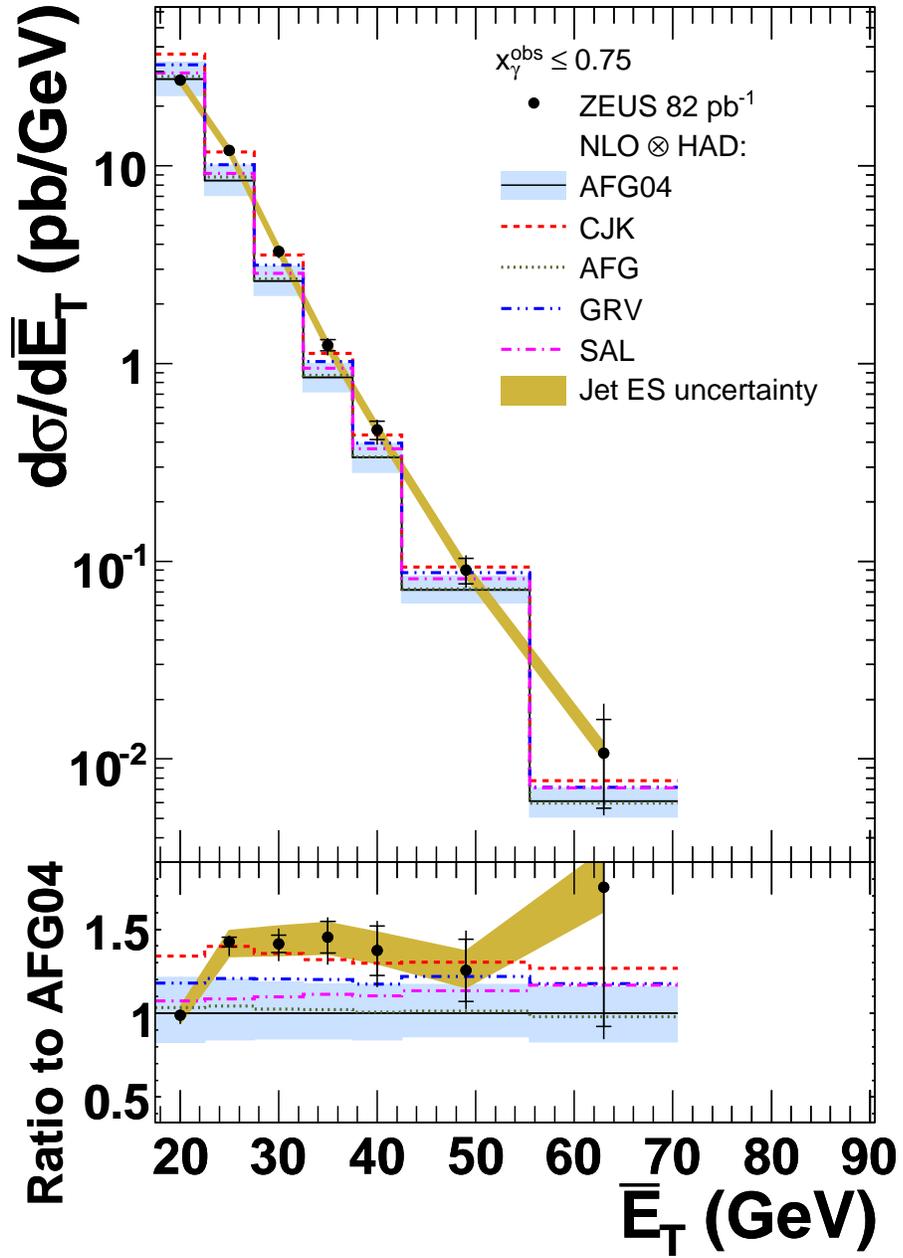,height=20cm}
\vspace{-1cm}
\caption{Measured cross-section for $d\sigma/d\bar{E_T}$ for \xgo $\leq$ 0.75. For further 
details, see the caption to Fig.~\ref{fig:photon_xgamma}.}
\label{fig:photon_et}
\end{center}
\end{figure}

\begin{figure}[htb]
\begin{center}
~\epsfig{file=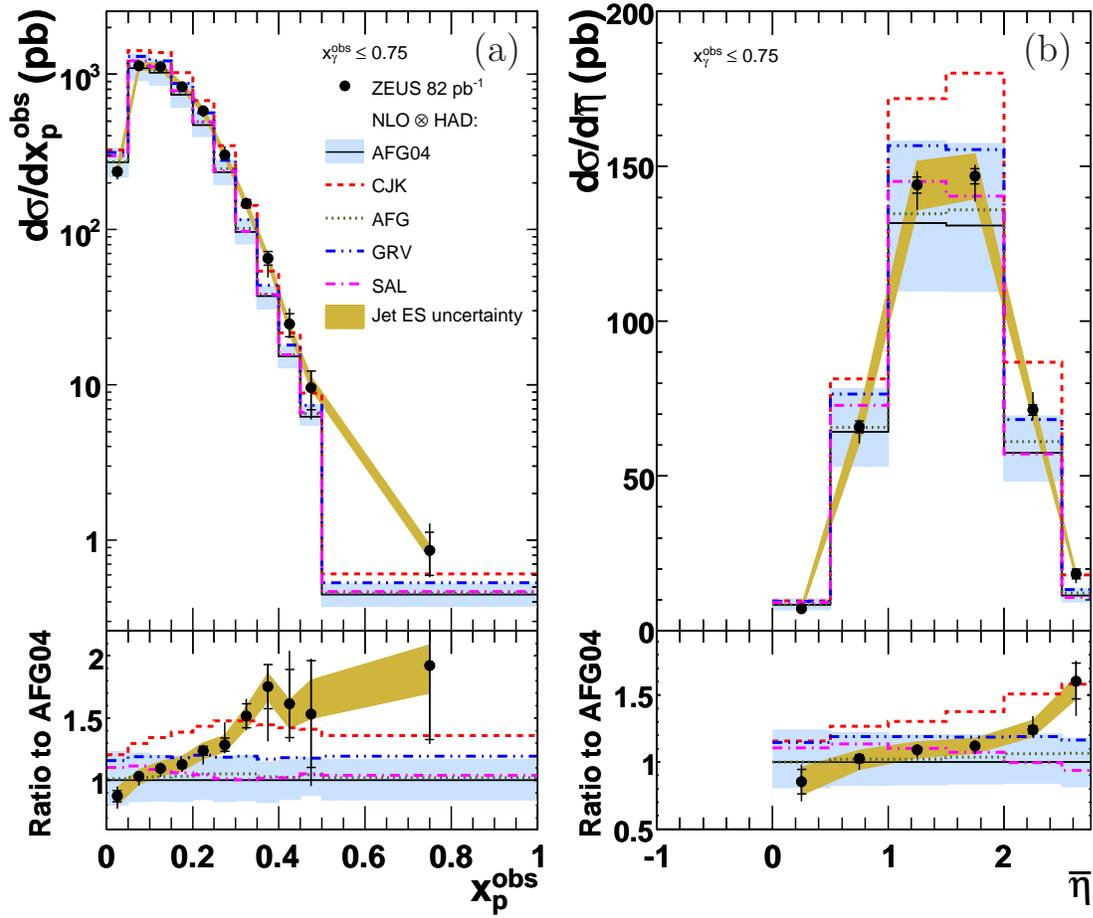,height=14.5cm}
\put(-240,358){\makebox(0,0)[tl]{\large (a)}}
\put(-32,358){\makebox(0,0)[tl]{\large (b)}}
\caption{Measured cross-section for (a) $d\sigma/dx_p^{\rm obs}$ and (b) $d\sigma/d\bar{\eta}$ 
both for \xgo $\leq$ 0.75. For further details, see the caption to Fig.~\ref{fig:photon_xgamma}.}
\label{fig:photon_others}
\end{center}
\end{figure}

As discussed in Section~\ref{sec-res1}, the measured cross sections show sensitivity 
to the choice of photon PDFs. This is to be expected due to the extension further 
forward in pseudorapidity compared to previous measurements. This was investigated 
further, with the results presented in Figs.~\ref{fig:photon_xgamma}--\ref{fig:photon_others}, 
where predictions with all five available parameterizations of the photon PDFs are compared 
to the data. In Table~\ref{tab:xgamma} and 
Fig.~\ref{fig:photon_xgamma} the cross-section $d\sigma/dx_\gamma^{\rm obs}$ 
is shown. At high $x_\gamma^{\rm obs}$, all predictions are similar, 
as expected since there is little sensitivity to the photon structure in this region. 
Towards low $x_\gamma^{\rm obs}$, the 
predictions differ by up to 70\%. The prediction from CJK deviates most from 
the other predictions and also from the data. The other predictions, although also 
exhibiting differences between each other of up to 25\%, give a qualitatively similar 
description of the data.

In Figs.~\ref{fig:photon_et} and~\ref{fig:photon_others}, the cross-sections 
$d\sigma/d\bar{E}_T$, $d\sigma/dx_p^{\rm obs}$ and $d\sigma/d\bar{\eta}$ are presented 
for $x_\gamma^{\rm obs} \leq 0.75$, as shown previously in 
Figs.~\ref{fig:gen_et},~\ref{fig:gen_xp} and~\ref{fig:gen_eta}, respectively, but 
here with additional predictions using different photon PDFs. For 
$d\sigma/d\bar{E}_T$, the prediction using CJK is much higher than the data in the 
first bin, but then agrees with the data for all subsequent bins. All photon PDFs 
have a similar shape, and none can reproduce the shape of the measured distribution.   
Apart from CJK, all PDFs are too low in the region $22.5 < \bar{E}_T < 37.5$~GeV. 
For the cross-section $d\sigma/dx_p^{\rm obs}$, no prediction gives a satisfactory 
description of the data. The prediction from CJK is generally above the data by 
20-30\%, but describes the shape of the cross section reasonably well. All other predictions 
give a poor description of the shape, with cross sections which fall too rapidly to 
high $x_p^{\rm obs}$. For $d\sigma/d\bar{\eta}$, the prediction from CJK again gives 
the best description of the shape of the data, although it is too high in normalization.

In summary, the data show a large sensitivity to the parameterization of the photon 
PDFs. The gluon PDF from CJK, in particular, differs from the others 
and this may give a hint of how to improve the photon PDFs. The data presented here 
should significantly improve the measurement of the gluon PDF of the photon, which is 
currently insufficiently constrained by the $F_2^\gamma$ data.

\section{Conclusions}
\label{sec-con}

Dijet cross sections in photoproduction have been measured at high $E_T^{\rm jet}$ 
and probe a wide range of \xgo \ and $x_p^{\rm obs}$. The kinematic region is 
$Q^2<1$\,GeV$^2$, 142 $<W_{\gamma p}<$ 293\,GeV, $E_T^{\rm jet1} >$ 20\,GeV, 
$E_T^{\rm jet2} >$ 15\,GeV and -1 $< \eta^{\rm jet1,2} <$ 3, with at least one jet 
lying in the range between $-1$ and $2.5$. 
In general, the data enriched in direct-photon events, at high \xgo, are well 
described by NLO QCD predictions. For the data enriched in resolved-photon events, 
at low \xgo, the data are less well described by NLO QCD predictions. Predictions 
using different parameterizations of the photon parton density functions give a large 
spread in the region measured, with no parton density function giving an 
adequate description of the data. Therefore the data have the potential to improve the 
constraints on the parton densities in the proton and photon and should be used in future 
fits. The cross section in the difference of azimuthal angle of the two jets is 
intrinsically sensitive to high-order QCD processes and the data are poorly described 
by NLO QCD, particularly at low \xgo. Therefore the data should be compared with new 
calculations of higher orders, or simulations thereof.

\section{Acknowledgments}

The strong support and encouragement of the DESY Directorate have been invaluable, 
and we are much indebted to the HERA machine group for their inventiveness and 
diligent efforts. The design, construction and installation of the ZEUS detector 
have been made possible by the ingenuity and dedicated efforts of many people from 
inside DESY and from the home institutes who are not listed as authors. Their 
contributions are acknowledged with great appreciation. We would also like to thank 
S. Frixione for help in using his calculation.
%------------------------------------------------------------------------------
%       Bibliography
%------------------------------------------------------------------------------
{
\def\bibname{\Large\bf References}
\def\refname{\Large\bf References}
\pagestyle{plain}
\ifzeusbst
  \bibliographystyle{./BiBTeX/bst/l4z_default}
\fi
\ifzdrftbst
  \bibliographystyle{./BiBTeX/bst/l4z_draft}
\fi
\ifzbstepj
  \bibliographystyle{./BiBTeX/bst/l4z_epj}
\fi
\ifzbstnp
  \bibliographystyle{./BiBTeX/bst/l4z_np}
\fi
\ifzbstpl
  \bibliographystyle{./BiBTeX/bst/l4z_pl}
\fi
{\raggedright
\bibliography{./BiBTeX/user/syn.bib,%
              ./BiBTeX/bib/l4z_articles.bib,%
              ./BiBTeX/bib/l4z_books.bib,%
              ./BiBTeX/bib/l4z_conferences.bib,%
              ./BiBTeX/bib/l4z_h1.bib,%
              ./BiBTeX/bib/l4z_misc.bib,%
              ./BiBTeX/bib/l4z_old.bib,%
              ./BiBTeX/bib/l4z_preprints.bib,%
              ./BiBTeX/bib/l4z_replaced.bib,%
              ./BiBTeX/bib/l4z_temporary.bib,%
              ./BiBTeX/bib/l4z_zeus.bib}}
}
\vfill\eject

%------------------------------------------------------------------------------
%       Tables
%------------------------------------------------------------------------------
%-------------------------------------------------------------------------------
%       An example table
%-------------------------------------------------------------------------------
\begin{table}[htb]
  \begin{center}
    \begin{tabular}{|c|c|c|c|}
\hline
Label & $x_\gamma^{\rm obs}$ cut    & $\eta^{\rm jet1,2}$ cuts & $E_T^{\rm jet1,2}$ cuts \\ \hline
``High-$x_\gamma^{\rm obs}$ 1''     & $x_\gamma^{\rm obs}>$ 0.75 & $0<\eta^{\rm jet1}<1$,   $2<\eta^{\rm jet2}<3$ & $E_T^{\rm jet1,2} > 25,15$~GeV \\ 
``High-$x_\gamma^{\rm obs}$ 2''     & $x_\gamma^{\rm obs}>$ 0.75 & $0<\eta^{\rm jet1}<1$,   $2<\eta^{\rm jet2}<3$ & $E_T^{\rm jet1,2} > 20,15$~GeV \\ 
``High-$x_\gamma^{\rm obs}$ 3''     & $x_\gamma^{\rm obs}>$ 0.75 & $1<\eta^{\rm jet1,2}<2$   & $E_T^{\rm jet1,2} > 30,15$~GeV \\ 
``High-$x_\gamma^{\rm obs}$ 4''     & $x_\gamma^{\rm obs}>$ 0.75 & $-1<\eta^{\rm jet1}<0$,  $0<\eta^{\rm jet2}<1$ & $E_T^{\rm jet1,2} > 20,15$~GeV \\ 
``Low-$x_\gamma^{\rm obs}$ 1''     & $x_\gamma^{\rm obs}<$ 0.75 & $2<\eta^{\rm jet1}<2.5$, $2<\eta^{\rm jet2}<3$ & $E_T^{\rm jet1,2} > 20, 15$~GeV \\ 
``Low-$x_\gamma^{\rm obs}$ 2''     & $x_\gamma^{\rm obs}<$ 0.75 & $1<\eta^{\rm jet1,2}<2$  & $E_T^{\rm jet1,2} > 25, 15$~GeV \\ 
``Low-$x_\gamma^{\rm obs}$ 3''     & $x_\gamma^{\rm obs}<$ 0.75 & $1<\eta^{\rm jet1}<2$,   $2<\eta^{\rm jet2}<3$ & $E_T^{\rm jet1,2} > 20,15$~GeV \\ 
``Low-$x_\gamma^{\rm obs}$ 4''     & $x_\gamma^{\rm obs}<$ 0.75 & $1<\eta^{\rm jet1}<2$,   $2<\eta^{\rm jet2}<3$ & $E_T^{\rm jet1,2} > 25,15$~GeV \\ 
\hline
    \end{tabular}
    \caption{Kinematic regions of the ``optimized'' cross sections.}
    \label{table:opt}
  \end{center}
\end{table}

\newpage

%%% s10056.tex

\begin{table}[hbt]
\begin{center}
\begin{tabular}{|c|cccccc|c|}
\hline
$\bar{E}_T$ bin (GeV) & $d\sigma/d\bar{E}_T$ & $\delta_{\rm stat}$ & $\delta_{\rm MC}$ & $\delta_{\rm syst}$ & $\delta_{\rm ES}$ & (pb/GeV) & $C_{\rm had}$ \\
\hline
17.5, 22.5 & 25.73  & $\pm$ 0.36  & $^{+0.66}_{-0.00}$ & $^{+0.41}_{-0.43}$ & $^{+1.03}_{-1.20}$      & & 0.955 $\pm$ 0.017 \\
22.5, 27.5 & 14.66  & $\pm$ 0.28  & $^{+0.00}_{-0.28}$ & $^{+0.42}_{-0.26}$ & $^{+0.60}_{-0.65}$      & & 0.931 $\pm$ 0.008 \\
27.5, 32.5 &  5.57  & $\pm$ 0.18  & $^{+0.09}_{-0.00}$ & $^{+0.14}_{-0.24}$ & $^{+0.30}_{-0.19}$      & & 0.937 $\pm$ 0.029 \\
32.5, 37.5 &  2.37  & $\pm$ 0.12  & $^{+0.00}_{-0.03}$ & $^{+0.15}_{-0.04}$ & $^{+0.11}_{-0.11}$      & & 0.927 $\pm$ 0.012 \\
37.5, 42.5 &  0.96  & $\pm$ 0.07  & $^{+0.02}_{-0.00}$ & $^{+0.06}_{-0.03}$ & $^{+0.07}_{-0.03}$      & & 0.907 $\pm$ 0.034 \\
42.5, 55.5 &  0.300 & $\pm$ 0.024 & $^{+0.000}_{-0.004}$ & $^{+0.004}_{-0.018}$ & $^{+0.016}_{-0.020}$& & 0.932 $\pm$ 0.044 \\
55.5, 70.5 &  0.046 & $\pm$ 0.009 & $^{+0.006}_{-0.000}$ & $^{+0.001}_{-0.003}$ & $^{+0.003}_{-0.003}$& & 0.926 $\pm$ 0.029 \\
70.5, 90.5 &  0.009 & $\pm$ 0.004 & $^{+0.001}_{-0.000}$ & $^{+0.001}_{-0.002}$ & $^{+0.000}_{-0.002}$& & 0.917 $\pm$ 0.085 \\
\hline
\end{tabular}
\caption{Measured cross-section $d\sigma/d\bar{E}_T$ for \xgo $>$ 0.75. The statistical, $\delta_{\rm stat}$, 
MC model, $\delta_{\rm MC}$, uncorrelated 
systematic, $\delta_{\rm syst}$, and jet energy scale, $\delta_{\rm ES}$, uncertainties are shown separately. 
The hadronization correction factor, $C_{\rm had}$, applied to the NLO QCD prediction is shown in the last 
column, where its uncertainty is half the spread between the values obtained using the {\sc Herwig} and 
{\sc Pythia} models.}
\label{tab:et-high}
\end{center}
\end{table}

%%% s10057.tex

\begin{table}[hbt]
\begin{center}
\begin{tabular}{|c|cccccc|c|}
\hline
$\bar{E}_T$ bin (GeV)  & $d\sigma/d\bar{E}_T$ & $\delta_{\rm stat}$ & $\delta_{\rm MC}$ & $\delta_{\rm syst}$ & $\delta_{\rm ES}$ & (pb/GeV) & $C_{\rm had}$ \\
\hline
17.5, 22.5 & 27.10 & $\pm$ 0.36 & $^{+0.49}_{-0.00}$ & $^{+0.18}_{-1.31}$ & $^{+1.45}_{-1.42}$       & & 1.082 $\pm$ 0.045 \\
22.5, 27.5 & 11.97 & $\pm$ 0.24 & $^{+0.07}_{-0.00}$ & $^{+0.21}_{-0.66}$ & $^{+0.56}_{-0.74}$       & & 1.047 $\pm$ 0.009 \\
27.5, 32.5 &  3.69 & $\pm$ 0.14 & $^{+0.17}_{-0.00}$ & $^{+0.10}_{-0.23}$ & $^{+0.27}_{-0.18}$       & & 1.057 $\pm$ 0.016 \\
32.5, 37.5 &  1.24 & $\pm$ 0.08 & $^{+0.03}_{-0.00}$ & $^{+0.06}_{-0.12}$ & $^{+0.07}_{-0.09}$       & & 1.004 $\pm$ 0.024 \\
37.5, 42.5 &  0.46 & $\pm$ 0.05 & $^{+0.03}_{-0.00}$ & $^{+0.01}_{-0.05}$ & $^{+0.04}_{-0.03}$       & & 1.069 $\pm$ 0.043 \\
42.5, 55.5 &  0.090 & $\pm$ 0.013 & $^{+0.005}_{-0.000}$ & $^{+0.009}_{-0.010}$ & $^{+0.008}_{-0.007}$ & & 1.019 $\pm$ 0.015 \\
55.5, 70.5 &  0.011 & $\pm$ 0.005 & $^{+0.004}_{-0.000}$ & $^{+0.006}_{-0.002}$  & $^{+0.001}_{-0.001}$ & & 0.974 $\pm$ 0.064 \\
\hline
\end{tabular}
\caption{Measured cross-section $d\sigma/d\bar{E}_T$ for \xgo $\leq$ 0.75. For further details, see the caption 
to Table~\ref{tab:et-high}.}
\label{tab:et-low}
\end{center}
\end{table}

%%% PostReadingIssues3/newtables/EtJet1HighXg.tex

\begin{table}[hbt]
\begin{center}
\begin{tabular}{|c|cccccc|c|}
\hline
$E_T^{\rm jet1}$ bin (GeV)  & $d\sigma/dE_T^{\rm jet1}$ & $\delta_{\rm stat}$ & $\delta_{\rm MC}$ & $\delta_{\rm syst}$ & $\delta_{\rm ES}$ & (pb/GeV) & $C_{\rm had}$ \\
\hline
20, 26 & 27.24 & $\pm$ 0.33  & $^{+0.18}_{-0.00}$ & $^{+0.56}_{-0.54}$ & $^{+1.05}_{-1.22}$ & & 0.957 $\pm$ 0.021 \\
26, 32 & 9.21  & $\pm$ 0.20  & $^{+0.17}_{-0.00}$ & $^{+0.21}_{-0.15}$ & $^{+0.49}_{-0.37}$ & & 0.920 $\pm$ 0.011 \\
32, 38 & 3.34  & $\pm$ 0.12  & $^{+0.00}_{-0.05}$ & $^{+0.16}_{-0.12}$ & $^{+0.14}_{-0.17}$ & & 0.916 $\pm$ 0.024 \\
38, 44 & 1.25  & $\pm$ 0.07  & $^{+0.03}_{-0.00}$ & $^{+0.15}_{-0.03}$ & $^{+0.07}_{-0.06}$ & & 0.943 $\pm$ 0.005 \\
44, 55 & 0.37  & $\pm$ 0.03  & $^{+0.00}_{-0.00}$ & $^{+0.01}_{-0.03}$ & $^{+0.02}_{-0.03}$ & & 0.921 $\pm$ 0.035 \\
55, 70 & 0.056 & $\pm$ 0.009 & $^{+0.008}_{-0.000}$ & $^{+0.004}_{-0.003}$ & $^{+0.007}_{-0.002}$ & & 0.889 $\pm$ 0.051 \\
70, 90 & 0.010 & $\pm$ 0.004 & $^{+0.004}_{-0.000}$ & $^{+0.004}_{-0.001}$ & $^{+0.002}_{-0.000}$ & & 0.85 $\pm$ 0.11 \\
\hline
\end{tabular}
\caption{Measured cross-section $d\sigma/dE_T^{\rm jet1}$ for \xgo $>$ 0.75. For further details, see the caption 
to Table~\ref{tab:et-high}.}
\label{tab:et1-high}
\end{center}
\end{table}

%%% PostReadingIssues3/newtables/EtJet1LowXg.tex

\begin{table}[hbt]
\begin{center}
\begin{tabular}{|c|cccccc|c|}
\hline
$E_T^{\rm jet1}$ bin (GeV)  & $d\sigma/dE_T^{\rm jet1}$ & $\delta_{\rm stat}$ & $\delta_{\rm MC}$ & $\delta_{\rm syst}$ & $\delta_{\rm ES}$ & (pb/GeV) & $C_{\rm had}$ \\
\hline
20, 26 & 25.59 & $\pm$ 0.31 & $^{+0.43}_{-0.00}$ & $^{+0.21}_{-1.33}$ & $^{+1.32}_{-1.34}$ & & 1.081 $\pm$ 0.043 \\
26, 32 & 8.11  & $\pm$ 0.18 & $^{+0.21}_{-0.00}$ & $^{+0.10}_{-0.41}$ & $^{+0.49}_{-0.47}$ & & 1.041 $\pm$ 0.015 \\
32, 38 & 2.39  & $\pm$ 0.10 & $^{+0.06}_{-0.00}$ & $^{+0.10}_{-0.17}$ & $^{+0.14}_{-0.15}$ & & 1.017 $\pm$ 0.025 \\
38, 44 & 0.72  & $\pm$ 0.05 & $^{+0.00}_{-0.01}$ & $^{+0.02}_{-0.05}$ & $^{+0.04}_{-0.05}$ & & 0.997 $\pm$ 0.006 \\
44, 55 & 0.18  & $\pm$ 0.02 & $^{+0.02}_{-0.00}$ & $^{+0.01}_{-0.02}$ & $^{+0.02}_{-0.01}$ & & 0.963 $\pm$ 0.027 \\
55, 70 & 0.018 & $\pm$ 0.006 & $^{+0.001}_{-0.000}$ & $^{+0.004}_{-0.003}$ & $^{+0.001}_{-0.002}$ & & 0.927 $\pm$ 0.033 \\
\hline
\end{tabular}
\caption{Measured cross-section $d\sigma/dE_T^{\rm jet1}$ for \xgo $\leq$ 0.75. For further details, see the caption 
to Table~\ref{tab:et-high}.}
\label{tab:et1-low}
\end{center}
\end{table}

%%% s10058.tex

\begin{table}[hbt]
\begin{center}
\begin{tabular}{|c|cccccc|c|}
\hline
$\bar{\eta}$ bin & $d\sigma/d\bar{\eta}$ & $\delta_{\rm stat}$ & $\delta_{\rm MC}$ & $\delta_{\rm syst}$ & $\delta_{\rm ES}$ & (pb) & $C_{\rm had}$ \\
\hline
-0.50, 0.00 &   4.8 & $\pm$ 1.2 & $^{+0.2}_{-0.0}$ & $^{+0.7}_{-1.4}$ & $^{+0.7}_{-1.6}$ & & 0.551 $\pm$ 0.037 \\
 0.00, 0.50 &  90.1 & $\pm$ 2.3 & $^{+5.1}_{-0.0}$ & $^{+4.0}_{-1.2}$ & $^{+6.8}_{-5.3}$ & & 0.892 $\pm$ 0.018 \\
 0.50, 1.00 & 177.8 & $\pm$ 2.9 & $^{+2.5}_{-0.0}$ & $^{+2.6}_{-3.6}$ & $^{+7.1}_{-8.9}$ & & 0.940 $\pm$ 0.001 \\
 1.00, 1.50 & 167.6 & $\pm$ 2.6 & $^{+0.0}_{-1.2}$ & $^{+6.5}_{-3.1}$ & $^{+6.6}_{-6.5}$ & & 0.952 $\pm$ 0.014 \\
 1.50, 2.00 &  59.0 & $\pm$ 1.5 & $^{+0.6}_{-0.0}$ & $^{+0.7}_{-0.6}$ & $^{+1.4}_{-1.5}$ & & 1.079 $\pm$ 0.035 \\
 2.00, 2.50 &   2.8 & $\pm$ 0.5 & $^{+0.0}_{-0.2}$ & $^{+0.1}_{-0.3}$ & $^{+0.0}_{-0.0}$ & & 1.062 $\pm$ 0.064 \\
\hline
\end{tabular}
\caption{Measured cross-section $d\sigma/d\bar{\eta}$ for \xgo $>$ 0.75. For further details, see the caption 
to Table~\ref{tab:et-high}.}
\label{tab:eta-high}
\end{center}
\end{table}

%%% s10059.tex

\begin{table}[hbt]
\begin{center}
\begin{tabular}{|c|cccccc|c|}
\hline
$\bar{\eta}$ bin & $d\sigma/d\bar{\eta}$ & $\delta_{\rm stat}$ & $\delta_{\rm MC}$ & $\delta_{\rm syst}$ & $\delta_{\rm ES}$ & (pb) & $C_{\rm had}$ \\
\hline
0.00, 0.50 &   7.2 & $\pm$ 0.8 & $^{+0.0}_{-0.1}$ & $^{+0.7}_{-0.9}$ & $^{+0.9}_{-0.8}$ & & 1.052 $\pm$ 0.080 \\
0.50, 1.00 &  65.9 & $\pm$ 1.9 & $^{+0.0}_{-0.0}$ & $^{+1.5}_{-5.1}$ & $^{+4.1}_{-5.1}$ & & 1.074 $\pm$ 0.054 \\
1.00, 1.50 & 144.0 & $\pm$ 2.6 & $^{+3.2}_{-0.0}$ & $^{+1.7}_{-7.6}$ & $^{+7.6}_{-8.1}$ & & 1.080 $\pm$ 0.021 \\
1.50, 2.00 & 146.8 & $\pm$ 2.4 & $^{+1.6}_{-0.0}$ & $^{+2.2}_{-7.8}$ & $^{+7.2}_{-7.2}$ & & 1.063 $\pm$ 0.019 \\
2.00, 2.50 &  71.3 & $\pm$ 1.7 & $^{+5.1}_{-0.0}$ & $^{+2.2}_{-2.5}$ & $^{+4.0}_{-2.9}$ & & 1.062 $\pm$ 0.022 \\
2.50, 2.75 &  18.4 & $\pm$ 1.5 & $^{+0.7}_{-0.0}$ & $^{+0.3}_{-2.6}$ & $^{+0.4}_{-1.5}$ & & 1.066 $\pm$ 0.002 \\
\hline
\end{tabular}
\caption{Measured cross-section $d\sigma/d\bar{\eta}$ for \xgo $\leq$ 0.75. For further details, see the caption 
to Table~\ref{tab:et-high}.}
\label{tab:eta-low}
\end{center}
\end{table}

\newpage

%%% s10052.tex

\begin{table}[hbt]
\begin{center}
\begin{tabular}{|c|cccccc|c|}
\hline
$x_p^{\rm obs}$ bin & $d\sigma/dx_p^{\rm obs}$ & $\delta_{\rm stat}$ & $\delta_{\rm MC}$ & $\delta_{\rm syst}$ & $\delta_{\rm ES}$ & (pb) & $C_{\rm had}$ \\
\hline
0.00, 0.05 & 1260 &    $\pm$ 26   & $^{+57}_{-0}$ & $^{+21}_{-23}$ & $^{+69}_{-72}$ & & 0.902 $\pm$ 0.025 \\
0.05, 0.10 & 1960 &    $\pm$ 30   & $^{+7}_{-0}$ & $^{+35}_{-48}$ & $^{+81}_{-82}$ & & 0.932 $\pm$ 0.007 \\
0.10, 0.15 &  925 &    $\pm$ 20   & $^{+0}_{-1}$ & $^{+60}_{-12}$  & $^{+27}_{-41}$ & & 0.996 $\pm$ 0.024 \\
0.15, 0.20 &  468 &    $\pm$ 15   & $^{+0}_{-9}$ & $^{+13}_{-7}$ & $^{+24}_{-17}$ & & 0.999 $\pm$ 0.015 \\
0.20, 0.25 &  220 &    $\pm$ 11   & $^{+0}_{-4}$ & $^{+12}_{-5}$ & $^{+6}_{-9}$ & & 0.982 $\pm$ 0.012 \\
0.25, 0.30 &  104.9 &  $\pm$ 8.4  & $^{+0.0}_{-1.3}$ & $^{+2.9}_{-10.8}$ & $^{+5.1}_{-4.1}$ & & 0.963 $\pm$ 0.015 \\
0.30, 0.35 &   45.0 &  $\pm$ 5.6  & $^{+1.5}_{-0.0}$ & $^{+3.4}_{-1.0}$ & $^{+2.4}_{-1.2}$ & & 1.063 $\pm$ 0.023 \\
0.35, 0.40 &   23.2 &  $\pm$ 4.1  & $^{+0.0}_{-0.9}$ & $^{+0.5}_{-0.9}$ & $^{+0.6}_{-1.6}$ & & 1.027 $\pm$ 0.008 \\
0.40, 0.45 &    8.7 &  $\pm$ 2.4  & $^{+0.9}_{-0.0}$ & $^{+4.0}_{-0.5}$ & $^{+1.0}_{-0.1}$ & & 1.010 $\pm$ 0.020 \\
0.45, 0.50 &    3.2 &  $\pm$ 1.4  & $^{+0.0}_{-0.3}$ & $^{+2.5}_{-1.0}$ & $^{+0.2}_{-0.2}$ & & 1.006 $\pm$ 0.016 \\
0.50, 1.00 &    0.40 & $\pm$ 0.17 & $^{+0.08}_{-0.00}$ & $^{+0.08}_{-0.21}$ & $^{+0.06}_{-0.01}$ & & 0.987 $\pm$ 0.018 \\
\hline
\end{tabular}
\caption{Measured cross-section $d\sigma/dx_p^{\rm obs}$ for \xgo $>$ 0.75. For further details, see the caption 
to Table~\ref{tab:et-high}.}
\label{tab:xp-high}
\end{center}
\end{table}

%%% s10053.tex

\begin{table}[hbt]
\begin{center}
\begin{tabular}{|c|cccccc|c|}
\hline
$x_p^{\rm obs}$ bin & $d\sigma/dx_p^{\rm obs}$ & $\delta_{\rm stat}$ & $\delta_{\rm MC}$ & $\delta_{\rm syst}$ & $\delta_{\rm ES}$ & (pb) & $C_{\rm had}$ \\
\hline
0.00, 0.05 &  236    & $\pm$ 12   & $^{+2}_{-0}$ & $^{+17}_{-24}$ & $^{+18}_{-19}$ & & 1.103 $\pm$ 0.092 \\
0.05, 0.10 & 1131    & $\pm$ 24   & $^{+0}_{-0}$ & $^{+19}_{-76}$ & $^{+55}_{-70}$ & & 1.063 $\pm$ 0.046 \\
0.10, 0.15 & 1120    & $\pm$ 22   & $^{+19}_{-0}$ & $^{+37}_{-63}$ & $^{+56}_{-61}$ & & 1.086 $\pm$ 0.022 \\
0.15, 0.20 &  829    & $\pm$ 19   & $^{+12}_{-0}$ & $^{+7}_{-37}$ & $^{+46}_{-37}$ & & 1.074 $\pm$ 0.001 \\
0.20, 0.25 &  581    & $\pm$ 17   & $^{+14}_{-0}$ & $^{+5}_{-49}$ & $^{+31}_{-30}$ & & 1.053 $\pm$ 0.001 \\
0.25, 0.30 &  302    & $\pm$ 12   & $^{+31}_{-0}$ & $^{+25}_{-10}$ & $^{+17}_{-13}$ & & 1.052 $\pm$ 0.052 \\
0.30, 0.35 &  146.8  & $\pm$ 9.4  & $^{+8.3}_{-0.0}$ & $^{+4.2}_{-6.2}$ & $^{+7.0}_{-9.7}$ & &    1.052 $\pm$ 0.014 \\
0.35, 0.40 &   65.5  & $\pm$ 6.6  & $^{+0.0}_{-0.3}$ & $^{+0.6}_{-15.0}$ & $^{+3.9}_{-4.2}$ & & 1.041 $\pm$ 0.008 \\
0.40, 0.45 &   24.6  & $\pm$ 4.2  & $^{+1.1}_{-0.0}$ & $^{+4.8}_{-2.2}$ & $^{+0.4}_{-3.0}$ & &  1.036 $\pm$ 0.004 \\
0.45, 0.50 &    9.6  & $\pm$ 2.7  & $^{+0.0}_{-0.7}$ & $^{+0.7}_{-2.3}$ & $^{+1.7}_{-0.2}$ & &  1.020 $\pm$ 0.005 \\
0.50, 1.00 &    0.86 & $\pm$ 0.27 & $^{+0.09}_{-0.00}$ & $^{+0.32}_{-0.09}$ & $^{+0.07}_{-0.10}$ & & 1.012 $\pm$ 0.034 \\
\hline
\end{tabular}
\caption{Measured cross-section $d\sigma/dx_p^{\rm obs}$ for \xgo $\leq$ 0.75. For further details, see the caption 
to Table~\ref{tab:et-high}.}
\label{tab:xp-low}
\end{center}
\end{table}

\newpage

%%% s10081.tex

\begin{table}[hbt]
\begin{center}
\begin{tabular}{|c|cccccc|c|}
\hline
$|\Delta\phi^{\rm jj}|$ bin & $d\sigma/d|\Delta\phi^{\rm jj}|$ & $\delta_{\rm stat}$ & $\delta_{\rm MC}$ & $\delta_{\rm syst}$ & $\delta_{\rm ES}$ & (pb/rad) & $C_{\rm had}$ \\
\hline
1.83, 2.09 &   1.7 & $\pm$ 0.5 & $^{+0.1}_{-0.0}$ & $^{+0.2}_{-0.5}$ & $^{+0.1}_{-0.2}$ & & 0.65 $\pm$ 0.11 \\
2.09, 2.36 &   7.8 & $\pm$ 1.0 & $^{+0.0}_{-0.0}$ & $^{+1.2}_{-0.6}$ & $^{+0.6}_{-0.6}$ & & 0.729 $\pm$ 0.059 \\
2.36, 2.62 &  36.1 & $\pm$ 2.2 & $^{+0.2}_{-0.0}$ & $^{+1.6}_{-1.7}$ & $^{+2.1}_{-1.8}$ & & 0.826 $\pm$ 0.013 \\
2.62, 2.88 & 132.9 & $\pm$ 3.9 & $^{+5.8}_{-0.0}$ & $^{+5.9}_{-2.7}$ & $^{+6.6}_{-8.3}$ & & 0.868 $\pm$ 0.008 \\
2.88, 3.14 & 779.1 & $\pm$ 8.1 & $^{+4.0}_{-0.0}$ & $^{+15.0}_{-13.3}$ & $^{+31.8}_{-33.6}$ & & 0.984 $\pm$ 0.015 \\
\hline
\end{tabular}
\caption{Measured cross-section $d\sigma/d|\Delta\phi^{\rm jj}|$ for \xgo $>$ 0.75. For further details, see the caption 
to Table~\ref{tab:et-high}.}
\label{tab:phi-high}
\end{center}
\end{table}

%%% s10082.tex

\begin{table}[hbt]
\begin{center}
\begin{tabular}{|c|cccccc|c|}
\hline
$|\Delta\phi^{\rm jj}|$ bin & $d\sigma/d|\Delta\phi^{\rm jj}|$ & $\delta_{\rm stat}$ & $\delta_{\rm MC}$ & $\delta_{\rm syst}$ & $\delta_{\rm ES}$ & (pb/rad) & $C_{\rm had}$ \\
\hline
0.00, 1.57 &   0.26 & $\pm$ 0.07 & $^{+0.05}_{-0.00}$ & $^{+0.02}_{-0.02}$ & $^{+0.04}_{-0.02}$ & & 0.84 $\pm$ 0.15 \\
1.57, 1.83 &   2.9  & $\pm$ 0.6  & $^{+0.3}_{-0.0}$ & $^{+0.6}_{-0.1}$ & $^{+0.1}_{-0.3}$ & & 0.869 $\pm$ 0.083 \\
1.83, 2.09 &   6.6  & $\pm$ 0.8  & $^{+0.2}_{-0.0}$ & $^{+0.4}_{-0.2}$ & $^{+0.3}_{-0.6}$ & & 0.910 $\pm$ 0.031 \\
2.09, 2.36 &  28.2  & $\pm$ 1.7  & $^{+0.0}_{-0.5}$ & $^{+0.6}_{-2.3}$ & $^{+2.4}_{-1.3}$ & & 0.959 $\pm$ 0.004 \\
2.36, 2.62 &  78.4  & $\pm$ 2.8  & $^{+1.2}_{-0.0}$ & $^{+3.5}_{-1.0}$ & $^{+4.3}_{-5.3}$ & & 0.988 $\pm$ 0.006 \\
2.62, 2.88 & 203.2  & $\pm$ 4.5  & $^{+0.0}_{-1.1}$ & $^{+0.6}_{-8.6}$ & $^{+10.4}_{-13.4}$ & & 1.006 $\pm$ 0.015 \\
2.88, 3.14 & 528.6  & $\pm$ 6.7  & $^{+16.5}_{-0.0}$ & $^{+6.0}_{-36.5}$ & $^{+28.1}_{-26.4}$ & & 1.069 $\pm$ 0.020 \\
\hline
\end{tabular}
\caption{Measured cross-section $d\sigma/d|\Delta\phi^{\rm jj}|$ for \xgo $\leq$ 0.75. For further details, see the caption 
to Table~\ref{tab:et-high}.}
\label{tab:phi-low}
\end{center}
\end{table}

\newpage

%%% s10065.tex

\begin{table}[hbt]
\begin{center}
\begin{tabular}{|c|cccccc|c|}
\hline
$x_p^{\rm obs}$ bin & $d\sigma/dx_p^{\rm obs}$ & $\delta_{\rm stat}$ & $\delta_{\rm MC}$ & $\delta_{\rm syst}$ & $\delta_{\rm ES}$ & (pb) & $C_{\rm had}$ \\
\hline
0.1, 0.2 & 80.9 & $\pm$ 4.2 & $^{+0.0}_{-3.4}$ & $^{+3.8}_{-6.1}$ & $^{+3.8}_{-3.4}$ & & 0.957 $\pm$ 0.010 \\
0.2, 0.3 & 51.6 & $\pm$ 3.5 & $^{+0.0}_{-1.0}$ & $^{+3.1}_{-2.0}$ & $^{+2.4}_{-2.1}$ & & 0.974 $\pm$ 0.059 \\
0.3, 0.4 & 12.6 & $\pm$ 2.1 & $^{+0.0}_{-0.0}$ & $^{+1.0}_{-0.9}$ & $^{+0.6}_{-0.9}$ & & 0.962 $\pm$ 0.010 \\
0.4, 0.5 &  2.1 & $\pm$ 1.0 & $^{+1.0}_{-0.0}$ & $^{+1.0}_{-0.3}$ & $^{+0.2}_{-0.1}$ & & 0.953 $\pm$ 0.024 \\
\hline
\end{tabular}
\caption{Measured cross-section $d\sigma/dx_p^{\rm obs}$ for \xgo $>$ 0.75 (``High-\xgo \ 1''). For further details, see the caption 
to Table~\ref{tab:et-high}.}
\label{tab:optd}
\end{center}
\end{table}

%%% s10066.tex

\begin{table}[hbt]
\begin{center}
\begin{tabular}{|c|cccccc|c|}
\hline
$x_p^{\rm obs}$ bin & $d\sigma/dx_p^{\rm obs}$ & $\delta_{\rm stat}$ & $\delta_{\rm MC}$ & $\delta_{\rm syst}$ & $\delta_{\rm ES}$ & (pb) & $C_{\rm had}$ \\
\hline
0.0, 0.1 &  10.1 & $\pm$ 1.6 & $^{+0.1}_{-0.0}$ & $^{+0.6}_{-0.5}$ & $^{+0.7}_{-0.2}$ & &   0.961 $\pm$ 0.037 \\
0.1, 0.2 & 238.9 & $\pm$ 7.1 & $^{+0.0}_{-5.2}$ & $^{+15.0}_{-6.8}$ & $^{+9.7}_{-10.8}$ & & 1.006 $\pm$ 0.021 \\
0.2, 0.3 &  77.0 & $\pm$ 4.5 & $^{+0.0}_{-2.4}$ & $^{+6.7}_{-1.9}$ & $^{+3.6}_{-2.7}$ & &   1.005 $\pm$ 0.026 \\
0.3, 0.4 &  12.6 & $\pm$ 2.1 & $^{+0.0}_{-0.0}$ & $^{+0.9}_{-0.9}$ & $^{+0.6}_{-0.9}$ & &   0.964 $\pm$ 0.009 \\
0.4, 0.5 &   2.1 & $\pm$ 1.0 & $^{+1.0}_{-0.0}$ & $^{+1.0}_{-0.3}$ & $^{+0.2}_{-0.1}$ & &   0.953 $\pm$ 0.024 \\
\hline
\end{tabular}
\caption{Measured cross-section $d\sigma/dx_p^{\rm obs}$ for \xgo $>$ 0.75 (``High-\xgo \ 2''). For further details, see the caption 
to Table~\ref{tab:et-high}.}
\label{tab:opte}
\end{center}
\end{table}

%%% s10067.tex

\begin{table}[hbt]
\begin{center}
\begin{tabular}{|c|cccccc|c|}
\hline
$x_p^{\rm obs}$ bin & $d\sigma/dx_p^{\rm obs}$ & $\delta_{\rm stat}$ & $\delta_{\rm MC}$ & $\delta_{\rm syst}$ & $\delta_{\rm ES}$ & (pb) & $C_{\rm had}$ \\
\hline
0.0, 0.1 &  2.1 & $\pm$ 0.8 & $^{+0.4}_{-0.0}$ & $^{+1.4}_{-0.1}$ & $^{+0.1}_{-0.1}$ & & 0.914 $\pm$ 0.014 \\
0.1, 0.2 & 55.9 & $\pm$ 3.5 & $^{+0.1}_{-0.0}$ & $^{+1.2}_{-2.7}$ & $^{+2.3}_{-1.4}$ & & 0.974 $\pm$ 0.006 \\
0.2, 0.3 & 20.5 & $\pm$ 2.1 & $^{+0.9}_{-0.0}$ & $^{+0.3}_{-3.0}$ & $^{+0.7}_{-0.8}$ & & 0.988 $\pm$ 0.011 \\
0.3, 0.4 &  2.4 & $\pm$ 0.7 & $^{+0.0}_{-0.0}$ & $^{+0.1}_{-0.4}$ & $^{+0.1}_{-0.1}$ & & 1.007 $\pm$ 0.046 \\
\hline
\end{tabular}
\caption{Measured cross-section $d\sigma/dx_p^{\rm obs}$ for \xgo $>$ 0.75 (``High-\xgo \ 3''). For further details, see the caption 
to Table~\ref{tab:et-high}.}
\label{tab:optf}
\end{center}
\end{table}

%%% s10068.tex

\begin{table}[hbt]
\begin{center}
\begin{tabular}{|c|cccccc|c|}
\hline
$x_p^{\rm obs}$ bin & $d\sigma/dx_p^{\rm obs}$ & $\delta_{\rm stat}$ & $\delta_{\rm MC}$ & $\delta_{\rm syst}$ & $\delta_{\rm ES}$ & (pb) & $C_{\rm had}$ \\
\hline
0.0, 0.1 & 198.0 & $\pm$ 8.8 & $^{+10.9}_{-0.0}$ & $^{+2.9}_{-2.3}$ & $^{+18.7}_{-16.0}$ & & 0.832 $\pm$ 0.017 \\
\hline
\end{tabular}
\caption{Measured cross-section $d\sigma/dx_p^{\rm obs}$ for \xgo $>$ 0.75 (``High-\xgo \ 4''). For further details, see the caption 
to Table~\ref{tab:et-high}.}
\label{tab:opth}
\end{center}
\end{table}

%%% s10061.tex

\begin{table}[hbt]
\begin{center}
\begin{tabular}{|c|cccccc|c|}
\hline
$x_p^{\rm obs}$ bin & $d\sigma/dx_p^{\rm obs}$ & $\delta_{\rm stat}$ & $\delta_{\rm MC}$ & $\delta_{\rm syst}$ & $\delta_{\rm ES}$ & (pb) & $C_{\rm had}$ \\
\hline
0.1, 0.2 & 15.0  & $\pm$ 2.0  & $^{+0.8}_{-0.0}$ & $^{+2.2}_{-0.5}$ & $^{+0.5}_{-0.3}$  & & 1.004 $\pm$ 0.099 \\
0.2, 0.3 & 89.4  & $\pm$ 4.6  & $^{+13.4}_{-0.0}$ & $^{+1.5}_{-4.1}$ & $^{+4.3}_{-3.9}$ & & 1.030 $\pm$ 0.003 \\
0.3, 0.4 & 46.7  & $\pm$ 3.8  & $^{+2.3}_{-0.0}$ & $^{+0.4}_{-4.3}$ & $^{+1.8}_{-3.3}$  & & 1.070 $\pm$ 0.090 \\
0.4, 0.5 &  7.0  & $\pm$ 1.5  & $^{+0.4}_{-0.0}$ & $^{+0.2}_{-0.6}$ & $^{+0.1}_{-0.9}$  & & 0.960 $\pm$ 0.083 \\
0.5, 1.0 &  0.48 & $\pm$ 0.20 & $^{+0.00}_{-0.04}$ & $^{+0.04}_{-0.09}$ & $^{+0.03}_{-0.05}$ & & 1.024 $\pm$ 0.027 \\
\hline
\end{tabular}
\caption{Measured cross-section $d\sigma/dx_p^{\rm obs}$ for \xgo $\leq$ 0.75 (``Low-\xgo \ 1''). For further details, see the caption 
to Table~\ref{tab:et-high}.}
\label{tab:opta}
\end{center}
\end{table}

%%% s10062.tex

\begin{table}[hbt]
\begin{center}
\begin{tabular}{|c|cccccc|c|}
\hline
$x_p^{\rm obs}$ bin & $d\sigma/dx_p^{\rm obs}$ & $\delta_{\rm stat}$ & $\delta_{\rm MC}$ & $\delta_{\rm syst}$ & $\delta_{\rm ES}$ & (pb) & $C_{\rm had}$ \\
\hline
0.0, 0.1 &  19.5 & $\pm$ 2.3 & $^{+1.5}_{-0.0}$ & $^{+0.8}_{-3.0}$ & $^{+0.4}_{-1.8}$ & & 0.876 $\pm$ 0.076 \\
0.1, 0.2 & 117.6 & $\pm$ 5.0 & $^{+2.0}_{-0.0}$ & $^{+4.7}_{-9.7}$ & $^{+5.5}_{-5.3}$ & & 1.048 $\pm$ 0.014 \\
0.2, 0.3 &  12.6 & $\pm$ 1.7 & $^{+0.6}_{-0.0}$ & $^{+0.6}_{-1.9}$ & $^{+0.7}_{-0.7}$ & & 1.116 $\pm$ 0.085 \\
\hline
\end{tabular}
\caption{Measured cross-section $d\sigma/dx_p^{\rm obs}$ for \xgo $\leq$ 0.75 (``Low-\xgo \ 2''). For further details, see the caption 
to Table~\ref{tab:et-high}.}
\label{tab:optb}
\end{center}
\end{table}

%%% s10063.tex

\begin{table}[hbt]
\begin{center}
\begin{tabular}{|c|cccccc|c|}
\hline
$x_p^{\rm obs}$ bin & $d\sigma/dx_p^{\rm obs}$ & $\delta_{\rm stat}$ & $\delta_{\rm MC}$ & $\delta_{\rm syst}$ & $\delta_{\rm ES}$ & (pb) & $C_{\rm had}$ \\
\hline
0.1, 0.2 & 278.4  & $\pm$ 7.6  & $^{+4.2}_{-0.0}$ & $^{+4.6}_{-12.7}$ & $^{+13.5}_{-12.4}$ & & 1.087 $\pm$ 0.015 \\
0.2, 0.3 & 235.2  & $\pm$ 7.1  & $^{+10.3}_{-0.0}$ & $^{+2.1}_{-9.6}$ & $^{+12.2}_{-10.3}$ & & 1.077 $\pm$ 0.030 \\
0.3, 0.4 &  47.8  & $\pm$ 3.6  & $^{+0.7}_{-0.0}$ & $^{+0.8}_{-3.4}$ & $^{+2.8}_{-2.6}$    & & 0.999 $\pm$ 0.064 \\
0.4, 0.5 &   8.3  & $\pm$ 1.6  & $^{+0.0}_{-0.1}$ & $^{+1.7}_{-0.6}$  & $^{+0.7}_{-0.6}$    & & 1.037 $\pm$ 0.020 \\
0.5, 1.0 &   0.28 & $\pm$ 0.14 & $^{+0.15}_{-0.0}$ & $^{+0.19}_{-0.04}$ & $^{+0.07}_{-0.01}$ & & 1.003 $\pm$ 0.037 \\
\hline
\end{tabular}
\caption{Measured cross-section $d\sigma/dx_p^{\rm obs}$ for \xgo $\leq$ 0.75 (``Low-\xgo \ 3''). For further details, see the caption 
to Table~\ref{tab:et-high}.}
\label{tab:optc}
\end{center}
\end{table}

%%% s10064.tex

\begin{table}[hbt]
\begin{center}
\begin{tabular}{|c|cccccc|c|}
\hline
$x_p^{\rm obs}$ bin & $d\sigma/dx_p^{\rm obs}$ & $\delta_{\rm stat}$ & $\delta_{\rm MC}$ & $\delta_{\rm syst}$ & $\delta_{\rm ES}$ & (pb) & $C_{\rm had}$ \\
\hline
0.1, 0.2 & 71.3  & $\pm$ 4.1  & $^{+1.8}_{-0.0}$ & $^{+2.6}_{-4.6}$ & $^{+4.2}_{-3.4}$  & & 1.066 $\pm$ 0.052 \\
0.2, 0.3 & 120.4 & $\pm$ 5.0  & $^{+5.6}_{-0.0}$ & $^{+2.6}_{-6.3}$ & $^{+7.3}_{-4.6}$ & & 1.042 $\pm$ 0.021 \\
0.3, 0.4 & 45.0  & $\pm$ 3.4  & $^{+0.3}_{-0.0}$ & $^{+1.9}_{-3.3}$ & $^{+1.8}_{-3.2}$  & & 1.013 $\pm$ 0.059 \\
0.4, 0.5 & 8.3   & $\pm$ 1.6  & $^{+0.0}_{-0.1}$ & $^{+1.7}_{-0.6}$ & $^{+0.7}_{-0.6}$   & & 1.037 $\pm$ 0.020 \\
0.5, 1.0 & 0.28  & $\pm$ 0.14 & $^{+0.15}_{-0.00}$ & $^{+0.19}_{-0.04}$ & $^{+0.07}_{-0.01}$ & & 1.003 $\pm$ 0.037 \\
\hline
\end{tabular}
\caption{Measured cross-section $d\sigma/dx_p^{\rm obs}$ for \xgo $\leq$ 0.75 (``Low-\xgo \ 4''). For further details, see the caption 
to Table~\ref{tab:et-high}.}
\label{tab:optg}
\end{center}
\end{table}

\newpage
\pagebreak[4]

%%% s10051.tex

\begin{table}[hbt]
\begin{center}
\begin{tabular}{|c|cccccc|c|}
\hline
$x_\gamma^{\rm obs}$ bin & $d\sigma/dx_\gamma^{\rm obs}$ & $\delta_{\rm stat}$ & $\delta_{\rm MC}$ & $\delta_{\rm syst}$ & $\delta_{\rm ES}$ & (pb) & $C_{\rm had}$ \\
\hline
0.1, 0.2 &  169.5 & $\pm$  6.8 & $^{+19.6}_{-0.0}$ & $^{+2.3}_{-7.4}$ & $^{+14.7}_{-12.6}$  & & 1.081 $\pm$ 0.046 \\
0.2, 0.3 &  271.6 & $\pm$  8.0 & $^{+12.0}_{-0.0}$ & $^{+1.7}_{-8.2}$ & $^{+17.1}_{-14.3}$  & & 1.042 $\pm$ 0.056 \\
0.3, 0.4 &  325.7 & $\pm$  8.9 & $^{+0.3}_{-0.0}$ & $^{+2.5}_{-15.2}$ & $^{+16.2}_{-16.3}$  & & 1.065 $\pm$ 0.017 \\
0.4, 0.5 &  346.6 & $\pm$  9.3 & $^{+7.2}_{-0.0}$ & $^{+7.6}_{-15.3}$ & $^{+17.2}_{-19.0}$ & & 1.058 $\pm$ 0.023 \\
0.5, 0.6 &  385   & $\pm$ 10   & $^{+3}_{-0}$ & $^{+4}_{-21}$ & $^{+20}_{-19}$  & & 1.072 $\pm$ 0.016 \\
0.6, 0.7 &  458   & $\pm$ 11   & $^{+3}_{-0}$ & $^{+17}_{-30}$ & $^{+20}_{-24}$ & & 1.089 $\pm$ 0.028 \\
0.7, 0.8 &  557   & $\pm$ 12   & $^{+1}_{-0}$ & $^{+16}_{-55}$ & $^{+28}_{-29}$ & & 1.087 $\pm$ 0.011 \\
0.8, 1.0 & 1106   & $\pm$ 11   & $^{+15}_{-0}$ & $^{+32}_{-21}$ & $^{+47}_{-48}$ & & 0.940 $\pm$ 0.018 \\
\hline
\end{tabular}
\caption{Measured cross-section $d\sigma/dx_\gamma^{\rm obs}$. For further details, see the caption 
to Table~\ref{tab:et-high}.}
\label{tab:xgamma}
\end{center}
\end{table}

%------------------------------------------------------------------------------
%       Figures
%------------------------------------------------------------------------------
%-------------------------------------------------------------------------------
%       Results
%-------------------------------------------------------------------------------

\newpage

\newpage

\newpage

\newpage

\newpage

\newpage

\newpage

\newpage

\newpage

\newpage

\newpage

\newpage

\newpage

%
%       ... that's it
%
\end{document}